\documentclass[%
 reprint,
superscriptaddress,
nofootinbib,
 amsmath,amssymb,
 aps,
floatfix,
]{revtex4-2}

\usepackage{graphicx}
\usepackage{dcolumn}
\usepackage{bm}
\usepackage[dvipsnames]{xcolor}
\usepackage{ulem}
\usepackage[caption=false]{subfig}
\usepackage{lipsum}
\usepackage{tensor}
\usepackage[T1]{fontenc}

\usepackage[hidelinks]{hyperref}

\definecolor{TurkishBlue}{HTML}{144893}
\hypersetup{colorlinks=true,citecolor=TurkishBlue,linkcolor=TurkishBlue,urlcolor=TurkishBlue}

\usepackage [english]{babel}
\usepackage [autostyle, english = american]{csquotes}
\MakeOuterQuote{"}

\def\VEV#1{{\left\langle #1 \right \rangle}}

\newcommand{\dd}[0]{{\rm d}}

\newcommand{\td}[1]{{\tilde{#1}}}
\newcommand{\hOmega}{{\hat{\Omega}}}
\DeclareSymbolFont{toneitalic}{T1}{\familydefault}{m}{it}
\DeclareMathSymbol{\crpartial}{\mathord}{toneitalic}{"F0}
\newcommand{\apjl}{{Astrophys.\ J.\ Lett.}}

\begin{document}

\preprint{}

\title{Linear polarization of the stochastic gravitational-wave background \\ with pulsar timing arrays}

\author{Neha Anil Kumar}
\email{nanilku1@jhu.edu}
\affiliation{William H.\ Miller III Department of Physics and Astronomy, Johns Hopkins University, 3400 N.\ Charles St.,
Baltimore, Maryland 21218, USA}%

\author{Mesut \c{C}al{\i}\c{s}kan}
\affiliation{William H.\ Miller III Department of Physics and Astronomy, Johns Hopkins University, 3400 N.\ Charles St.,
Baltimore, Maryland 21218, USA}%

\author{Gabriela Sato-Polito}%
\affiliation{William H.\ Miller III Department of Physics and Astronomy, Johns Hopkins University, 3400 N.\ Charles St.,
Baltimore, Maryland 21218, USA}%
\affiliation{School of Natural Sciences, Institute for Advanced Study, Princeton, NJ 08540, United States}
\author{Marc Kamionkowski}
\affiliation{William H.\ Miller III Department of Physics and Astronomy, Johns Hopkins University, 3400 N.\ Charles St.,
Baltimore, Maryland 21218, USA}%

\author{Lingyuan Ji}
\affiliation{Department of Physics, University of California, Berkeley, 366 Physics North MC 7300, Berkeley, California 94720, USA}

%



\begin{abstract}
\noindent Pulsar-timing collaborations have recently reported evidence for the detection of an isotropic stochastic gravitational-wave background (SGWB) consistent with one sourced by a population of inspiraling supermassive black hole binaries. 
However, a certain degree of anisotropy and polarization may be present. 
Thus, the characterization of the energy density and polarization of the background at different angular scales is important. In this paper, we describe the signatures of linear polarization in the stochastic gravitational-wave background on the timing residuals obtained with pulsar-timing arrays. 
We expand the linear polarization map in terms of spin-weighted spherical harmonics and recast it into the $E$-mode (parity even) and $B$-mode (parity odd) basis.
We provide expressions for the minimum-variance estimators for the coefficients of that expansion and evaluate the smallest
detectable signal as a function of the signal-to-noise ratio with
which the isotropic GW signal is detected and the number of
pulsars in the survey. 
We evaluate the covariance between the estimators for the spherical-harmonic coefficients of the linear polarization $E$-modes and those for the intensity anisotropy. We also show that there is no covariance between the spherical-harmonic coefficients for the $B$-modes of the linear polarization and those for the circular polarization, even though both have the same parity. Under the simplifying assumptions of our harmonic analysis, we show that detection of the lowest-order linear-polarization anisotropies is only possible if the isotropic intensity contribution is measured with a signal-to-noise ratio $\geq 300$, with a pulsar-timing network comprised of more than $\sim 100$ pulsars. We also show that, in this limit, the lowest order anisotropies in the $E$-mode have a negligible covariance with corresponding anisotropies in the SGWB intensity.
Our approach results in simple, elegant, and easily evaluated expressions for the overlap reduction functions for linear polarization.
\end{abstract}

\maketitle

\section{\label{sec:Introduction}Introduction\protect}
Since the turn of the century, radio telescopes across the world have been timing a network of millisecond pulsars with the primary goal of gravitational-wave (GW) detection. Because the arrival times of the pulses are sensitive to gravitational radiation, the observed network of pulsars, called a pulsar timing array (PTA), behaves as a galactic-scale GW detector. With more than a decade of timing data, PTA collaborations \cite{NanoGrav:2023gor, EPTA:2023fyk, Reardon:2023gzh, Xu:2023wog} have finally reported evidence for the detection of a stochastic gravitational-wave background (SGWB) at $\sim$nHz frequencies, as expected from the inspirals of merging supermassive black hole binaries (SMBHBs)\cite{Rajagopal:1994zj, Jaffe:2002rt}.

The SGWB signal contained in the difference between the expected and actual arrival times of the pulses---called timing residuals---can be extracted by correlating the residuals observed from two different pulsars. If the origin of the background is assumed to be cosmological or is sourced by a large population of distant objects, the detected SGWB signal is expected to be isotropic. Under this assumption, the resulting pulsar correlations vary with the angle between the pulsars according to the so-called Hellings-Downs (HD) curve \cite{1983ApJ...265L..39H}. Excess noise in pulse arrival times with a frequency spectrum and amplitude consistent with a GW background \cite{NANOGrav:2020bcs, Chen:2021rqp, Goncharov:2021oub, Antoniadis:2022pcn} had been noted for several years, but only now have we seen evidence for detection of the HD curve.

Given this recent evidence for an SGWB in the PTA residuals, the next step will be to characterize the signal in more detail. One approach is to further characterize the isotropic signal, looking at additional information that can be gained from the HD correlation detection~\cite{Bernardo:2023mxc, Bernardo:2023bqx, Bernardo:2023pwt, Bernardo:2022xzl, NANOGrav:2023ygs} or considering cross-correlation prospects with astrometric observations of the SGWB~\cite{Qin:2018yhy}. Another approach is to move beyond the detection of isotropy. Although the assumption of statistical isotropy applies in certain regimes, some degree of anisotropy is expected. For example, if the signal is dominated by $\sim N$ sources, the anisotropy is expected to have an amplitude $\sim N^{-1/2}$ \cite{Ravi:2012bz, Cornish:2013aba, Sesana:2008mz, Gair:2014rwa} i.e., if the signal is dominated by a handful of nearby sources, the observed signal can be anisotropic. Furthermore, given that GWs from inspirals are most generally circularly and linearly polarized, the SGWB is also expected to exhibit circular- and linear-polarization anisotropies. 
Techniques have been developed~\cite{Taylor:2013esa, Hotinli:2019tpc, Ali-Haimoud:2020iyz, Ali-Haimoud:2020ozu, Pol:2022sjn}, and now applied \cite{Taylor:2015udp, NANOGrav:2023tcn}, to seek anisotropy in the intensity of the background with the observed pulse arrival times. Moreover, measurement prospects and detection methodologies for circular polarization have been explored~\cite{Kato:2015bye, Belgacem:2020nda, Sato-Polito:2021efu} with recent work also discussing the linearly polarized component~\cite{Chu:2021krj, Liu:2022skj}.

In this paper, we forecast the future ability to probe linear polarization in the SGWB with the timing residuals observed by PTAs, via the analysis of these residuals in harmonic space. 
To do so, we expand the direction dependence of the linear polarization of the SGWB in terms of spin-weighted spherical harmonics. Recasting the linear polarization into the more familiar $E$- and $B$-mode basis, we then derive minimum-variance estimators (and their variances) for the expansion coefficients in terms of the signal-to-noise ratio (SNR) of the isotropic HD detection.

We present this derivation in harmonic space, using the bipolar spherical harmonics (BiPoSH) formalism~\cite{Hajian:2003qq,Hajian:2005jh,Joshi:2009mj,Book:2011na}, following the analogous derivations for intensity anisotropies and circular polarization in Refs.~\cite{Hotinli:2019tpc,Belgacem:2020nda}.
Although this formalism is less familiar in the PTA literature, it facilitates straightforward analytic estimations of a PTA's sensitivity to linear polarization. Furthermore, it elucidates how this sensitivity scales with various PTA specifications, such as noise properties and the number of pulsars.

While some of our results reproduce those recently obtained in Refs.~\cite{Chu:2021krj,Liu:2022skj}, the methodology presented here may offer a more streamlined derivation. Not only do we go further by explicitly constructing the linear-polarization estimators and forecasting the sensitivity of PTAs to linear-polarization anisotropy,we also calculate the covariance between estimators for the $E$-mode linear polarization and those for the intensity anisotropies and show that there is no covariance between the $B$-mode linear polarization and the circular polarization, even though they have the same parity. Saliently, the derivation presented here also allows us to reveal a new expression for linear polarization ORFs that is compact, easy to compute, and applicable to any chosen coordinate system of observation.\footnote{A companion paper \cite{AnilKumar:2023yfw} generalizes this to intensity anisotropies and circular polarization and to the spin-1 GWs that may arise in alternative-gravity theories. Reference~\cite{Bernardo:2023jhs} also extends the formalism to describe correlations for a scalar SGWB, including the effects of the pulsar term. We hope that the clear derivation here acts as a supplement to the results summarized in the companion piece.}

The primary focus of this paper is the mathematical description of the angular correlations across timing residuals induced by a given SGWB angular linear-polarization pattern. 
Therefore, we treat the time-sequence data simply, assuming that it can be decomposed into frequency-domain Fourier modes of some fixed frequency $f$ (averaged over a given observational frequency bin). As with previous work on intensity anisotropy, the estimators and covariances derived in this paper are obtained by initially assuming measurements at a specific frequency $f$. The additional steps needed to connect the results presented here with the data are exactly the same as those for other analyses (see, for example, Refs.~\cite{Mingarelli:2013dsa, Sato-Polito:2021efu}). 
How the estimators for different frequencies are combined depends on the specific model for the linear polarization. 
For example, if the SGWB is a {\it bona fide} stochastic background and the linear polarization is assumed to be frequency independent, then the estimators for different frequencies can be added with inverse-quadrature weighting in the noise. If, on the other hand, the GW signal is due to a finite number of sources \cite{Sato-Polito:2023spo,Gardiner:2023zzr}, then the linear-polarization pattern may differ from one frequency band to the next, with higher polarization expected at higher frequencies.

This paper is organized as follows.
We begin in Sec.~\ref{sec:residuals} by reviewing the timing residual induced in a single pulsar by an SGWB. In Sec.~\ref{sec: Characterizing the SGWB}, we characterize the SGWB in terms of the angular dependence of its intensity, linear polarization, and circular polarization. 
We expand these angular dependencies in terms of (spin-weighted) spherical harmonics for intensity and circular polarization (linear polarization), introducing the expansion coefficients that will be the primary target for measurement.
Sec.~\ref{sec:biposh} contains a review of the BiPoSH formalism and establishes the motivation for subsequent calculations. 
Sec.~\ref{sec:harmonicspace} presents our main results in harmonic space. 
We first calculate the BiPoSH coefficients induced by any given linear-polarization expansion coefficient. We then derive the minimum-variance estimators for the linear-polarization expansion coefficients, and we derive expressions for their variances. 
We then discuss the covariance between the estimators for the linear polarization and those for the intensity anisotropy and circular polarization. 
In Section~\ref{sec:configuration}, we leverage the BiPoSH approach to derive new expressions for linear polarization ORFs that are far more compact and elegant than those in prior work. Finally, in Section \ref{sec:forecasts}, we calculate the sensitivity of PTAs to linear polarization, computed using the previously derived estimators.
We present our concluding remarks in Sec.~\ref{sec:conclusions}.

\section{Single Pulsar Timing Residuals}
\label{sec:residuals}

A gravitational wave that propagates between the Earth and a pulsar affects the arrival time of the pulses observed at Earth. 
The fractional change in the pulse frequency of the pulsar, as compared to the expected intrinsic frequency, induced by a metric perturbation $h_{ij}(t, \hat{\Omega}, {\bf{x}})$ from a GW propagating in the $\hat{\Omega}$ direction is given by~\cite{1978SvA....22...36S, 1979ApJ...234.1100D}:
\begin{equation}
 z_{a}(t|\hat{\Omega}) = \frac{1}{2}\frac{n_a^i n_a^j}{1 + \hat{\Omega} \cdot \hat{n}_a} \Delta h_{ij}\,,
 \label{eq: timing_resid_oneGW}
\end{equation}
where we have used Einstein summation conventions. In the above equation, the subscript of $a$ labels a given pulsar, and $\hat n_a$ is the location of that pulsar in the sky. Furthermore, we have defined $\Delta h_{ij} \equiv h_{ij}(t_e, \hat{\Omega}, {\bf x}_e) - h_{ij}(t_p, \hat{\Omega}, {\bf x}_a)$ as the difference between the metric perturbation arriving at the Solar System barycenter (located at ${\bf x}_e$), at time $t_e$, and at the pulsar (located at ${\bf x}_a$) at time $t_p$. To simplify the analysis, we choose a coordinate system with the origin located at the barycenter of the Solar System and the pulsar placed at a distance $L_a$ from it, such that $t_e = t$, ${\bf x}_e = 0$, $t_p = t - L_a$, and ${\bf x}_a = L_a\hat{n}_a$.

To extend Eq.~\eqref{eq: timing_resid_oneGW} to an SGWB, we express the GW amplitude at location ${\bf x}$, in the transverse-traceless gauge, as a superposition of waves of all frequencies $f$ coming from all directions as follows:
\begin{equation}
 h_{ij}(t, {\bf x}) = \int \dd f \int \dd ^2\hat{\Omega} \sum_{A = +, \times} \tilde{h}_A(f, \hat{\Omega})e_{ij}^A(\hat{\Omega})e^{-2\pi if(t - \hat{\Omega}\cdot{\bf x})},
 \label{eq: SGWB_real_amplitude}
\end{equation}
where index $A \in \{+, \times\}$ labels the polarization, and the Fourier amplitudes $\tilde{h}_A(f, \hat{\Omega})$ are complex functions that satisfy $\tilde{h}^*_A(f, \hat{\Omega}) = \tilde{h}_A(-f, \hat{\Omega})$.
The polarization tensors $e_{ij}^A(\hat{\Omega})$ are given by
\begin{eqnarray}
     e_{ij}^+(\hat{\Omega}) = \hat{p}_i\hat{p}_j- \hat{q}_i\hat{q}_j, \\
    e_{ij}^\times(\hat{\Omega}) = \hat{p}_i\hat{q}_j + \hat{q}_i\hat{p}_j,
\end{eqnarray}
where $\hat{p}$ and $\hat{q}$ are unit vectors perpendicular to the direction of propagation $\hat{\Omega}$. 

We can plug in the expression for the SGWB amplitude at location $\bf{x}$ from Eq.~\eqref{eq: SGWB_real_amplitude} into Eq.~\eqref{eq: timing_resid_oneGW} to obtain the total, fractional frequency shift from an SGWB:
\begin{multline}
 z_a(t) = \sum_{A = +, \times} \int \dd f \int \dd ^2\hat{\Omega}\ \tilde{h}_A(f, \hat{\Omega})F_a^A(\hat{\Omega})e^{-2\pi ift} \\ \times \left[1 - e^{2\pi ifL_a(1 + \hat{\Omega}\cdot\hat{n}_a)}\right]\,, 
\end{multline}
where $F_a^A(\hat{\Omega})$ is the antenna beam pattern, defined as
\begin{equation}
 F_a^A(\hat{\Omega}) = \frac{1}{2}\frac{n_a^i n_a^j e_{ij}^A(\hat{\Omega})}{(1 + \hat{\Omega}\cdot \hat{n}_a)}\,.
 \label{eq: beam_pattern_def}
\end{equation}
The explicit form of the antenna beam pattern, which depends on the assumed coordinate system and the location of the chosen pulsar (unit vector $\hat{n}_a$), has been calculated in App.~\ref{App: ORF Computational}. Therefore, the timing residual can be written in frequency space as:
\begin{equation}
 z_a(f) = \sum_{A}\int \dd ^2\hat{\Omega}\ \tilde{h}_A(f, \hat{\Omega})F_a^A(\hat{\Omega}) \left[1 - e^{2\pi ifL_a(1 + \hat{\Omega}\cdot\hat{n}_a)}\right].
 \label{eq: single_residual_overlap_h}
\end{equation}

Assuming that timing-residual information $z_a(f)$ has been collected for a large number of pulsars as a function of $f$, the collective dataset represents a map of $z(\hat{n})$ across the sky at each frequency. Therefore, the angular structure of these residuals can be expanded in terms of spherical harmonics as:
\begin{equation}
 z(f, \hat{n}) = \sum_{\ell=2}^{\infty}\sum_{m=-\ell}^{\ell} z_{\ell m}(f) Y_{\ell m}(\hat{n})\,,
 \label{eq: single_resid_exp}
\end{equation}
where the sum is only over $\ell \geq 2$ since a transverse-traceless SGWB only gives rise to timing-residual patterns with $\ell \geq 2$. The expansion coefficients can be obtained via the inverse transform:
\begin{equation}
 z_{\ell m}(f) = \int \dd ^2\hat{n}\ z(\hat{n})\ Y_{\ell m}^*(\hat{n})\,.
\end{equation}
It is vital to note that for linear polarization and uncorrelated GW modes, the time-domain Fourier amplitudes $z_a(f)$ can be made to be real through a suitable choice of phase, i.e., with the appropriate phase, $z(f, \hat{n})$ 
is a real map on the two-sphere. Therefore, in each frequency bin, we have the constraint $z_{\ell m}^* = (-1)^m z_{\ell, -m}$.

\section{Characterizing the SGWB}
\label{sec: Characterizing the SGWB}

The correlation induced across the timing residuals of pairs of pulsars heavily depends on the assumed SGWB statistics.
In this work, we assume a background that is not only Gaussian and stationary but also possibly polarized and anisotropic.
When applied to the Fourier decomposition presented in Eq.~\eqref{eq: SGWB_real_amplitude}, our assumptions characterize a background for which the Fourier amplitudes for different modes are statistically independent, and each is chosen from a random distribution with variance given by:
\begin{equation}
 \langle \tilde{h}_A^*(f, \hat{\Omega})\tilde{h}_{A'}(f', \hat{\Omega}')\rangle = \delta(f-f')\delta^2(\hat{\Omega},\hat{\Omega}')\mathcal{P}_{A, A'}(f, \hat{\Omega})\,.
 \label{eq: GW_strain_correlations}
\end{equation}
where the $\delta$'s above represent Dirac delta functions. Here, $\mathcal{P}_{A, A'}(f, \hat{\Omega})$ is the spectral density of the background, which depends on the polarization of the two GW strains and their propagation directions. This polarization tensor can be expressed as
\begin{equation}
 \mathcal{P}_{A, A'}(f, \hat{\Omega}) = \begin{pmatrix} I(f, \hat{\Omega}) + Q(f, \hat{\Omega}) & U(f, \hat{\Omega}) - i V(f, \hat{\Omega}) \\ U(f, \hat{\Omega}) + i V(f, \hat{\Omega}) & I(f, \hat{\Omega}) - Q(f, \hat{\Omega})\end{pmatrix}\,.
 \label{eq: polarization_corr_tensor}
\end{equation}
This is analogous to how the tensor is defined using Stokes parameters in standard electromagnetism. For the SGWB, we define the Stokes parameters in terms of the GW strain as follows \cite{Conneely:2018wis}:
\begin{equation}
\begin{aligned}
 I(f, \hat{\Omega}) &= \frac{1}{2}\langle|\tilde{h}_+|^2 + |\tilde{h}_\times|^2\rangle,\\
 Q(f, \hat{\Omega}) &= \frac{1}{2}\langle|\tilde{h}_+|^2 - |\tilde{h}_\times|^2\rangle,\\
 U(f, \hat{\Omega}) &= {\rm{Re}}\langle{\tilde{h}_+^*\tilde{h}_\times}\rangle = \frac{1}{2}\langle\tilde{h}_+^*\tilde{h}_\times + \tilde{h}_\times^*\tilde{h}_+ \rangle, \\
 V(f, \hat{\Omega}) &= {\rm{Im}}\langle{\tilde{h}_+^*\tilde{h}_\times}\rangle = \frac{1}{2i}\langle\tilde{h}_+^*\tilde{h}_\times - \tilde{h}_\times^*\tilde{h}_+ \rangle,
\label{eqn:stokes} 
\end{aligned}
\end{equation}
where $I(f,\hat{\Omega})$ is the intensity, $V(f,\hOmega)$ is the circular polarization, and $Q(f,\hat{\Omega})$ and $U(f,\hat{\Omega})$ characterize the linear polarization. 

For our analysis, in both harmonic and configuration space, we assume that the frequency and angular dependence of the relevant Stokes parameters are separable. This allows us to expand the angular dependence of the parameters in terms of spherical-harmonic functions
\begin{eqnarray}
 I(f,\hat{\Omega}) &= &I(f)\sum_{L = 0}^{\infty} \sum_{M = -L}^L c_{LM}^I Y_{LM}(\hat{\Omega}) ,
 \label{eqn:expansion_I} \\
 V(f,\hat{\Omega}) &= &I(f)\sum_{L = 0}^{\infty} \sum_{M = -L}^L c_{LM}^VY_{LM}(\hat{\Omega}) , 
 \label{eqn:expansion_V} \\
 P_{\pm}(f, \hat{\Omega}) &= &I(f)\sum_{L = 4}^\infty \sum_{M = -L}^L c_{LM}^{\pm} {_{\pm 4}Y_{LM}}(\hat{\Omega}),
 \label{eqn:expansion_QU} 
\end{eqnarray}
where $c_{LM}^X$ for $X\in\{I,\, V,\, +,\, -\}$ are the expansion coefficients, and we have defined the spin-4 fields $P_{\pm}(f,\hat{\Omega})\equiv(Q\pm iU)(f,\hat{\Omega})$, expanded in terms of spin-weighted spherical harmonics\footnote{Note that we use capitalized angular-momentum quantum number $LM$ for the expansions for the Stokes parameters and lowercase $\ell m$ for expansions of the timing residuals.} ${}_{\pm s} Y_{LM}(\hOmega)$ of spin $s=4$. 
Assuming that the linear polarization and intensity maps share the same frequency dependence, the prefactor $I(f)$ describes these maps averaged over the frequencies in a frequency band centered at $f$. 
With the parametrizations in Eq.~\eqref{eqn:expansion_QU}, if there is any frequency dependence in the polarization relative to the intensity, it is absorbed in the frequency dependence of $c_{LM}^\pm$. Moreover, we normalize $I(f)$ such that $c_{00}^I = 1$. The primary focus of this work will be the linear polarization Stokes parameters $Q(f,\hOmega)$ and $U(f,\hOmega)$. 
Therefore, from this point on, we set every instance of $V(f,\hOmega)$ to zero unless stated otherwise.

Since the intensity $I$ is a real quantity, $c_{LM}^I = (-1)^M (c_{L,-M}^{I})^*$, indicating there are $2L+1$ independent real numbers encoded in $c_{LM}^I$ for a given $L$. Although $Q$ and $U$ are also real, the polarization $P_+=(Q+iU)$ is a complex quantity. Therefore, there is no analogous relation between $c_{LM}^+$ and its complex conjugate. However, we can still place a similar constraint by noting that $P_- = (Q-iU)$ is the complex conjugate of $P_+$. This requirement, applied to their respective expansions, allows us to place the constraint
\begin{equation}
 c_{LM}^+ = (-1)^M (c_{L, -M}^-)^*\,,
 \label{eq: clm_lin_relation}
\end{equation}
which follows from the relation ${_4}Y_{LM}^* = (-1)^M\,{_{-4}}Y_{L,-M}$. Therefore, given a measurement of the real and imaginary parts of $c_{LM}^+$, which amount to a total of $2(2L + 1)$ real numbers at fixed $L$, we can reconstruct both $Q$ and $U$, as expected. 

Since $Q$ and $U$ are not coordinate invariant, it is sometimes more convenient to define the scalar $E$ and $B$ fields using spin-raising and -lowering operators, resulting in the following expansions:
\begin{eqnarray}
 E(f, \hat{\Omega}) = I(f)\sum_{LM}c_{LM}^E Y_{LM}\,, && \\ 
 B(f, \hat{\Omega}) = I(f)\sum_{LM} c_{LM}^B Y_{LM}\,,
\end{eqnarray}
where the expansion coefficients can be defined in terms of $c_{L M}^{\pm}$ as:
\begin{align}
 c_{LM}^E = \frac{1}{2}\left(c_{LM}^+ + c_{LM}^-\right), && c_{LM}^B = -\frac{i}{2}\left(c_{LM}^+ - c_{LM}^-\right)\,.
 \label{eq: plus_minus_to_EB}
\end{align}
The functions $E(f,\hOmega)$ and $B(f,\hOmega)$ are real functions on the sphere, and so the expansion coefficients satisfy $(c_{LM}^{E})^*= (-1)^M c_{L,-M}^E$ and $(c_{LM}^{B})^*= (-1)^M c_{L,-M}^B$. The $E$-modes have even parity (as does the intensity), while the $B$-modes have odd parity (similar to circular polarization). 

For a linearly polarized SGWB, the Stokes parameters $I,\, Q$ and $U$ are subject to the constraint $|Q|^2 + |U|^2 \leq I^2$. Given the relations described in Eq.~\eqref{eq: plus_minus_to_EB} and the expansion in Eq.~\eqref{eqn:expansion_QU}, $Q$ and $U$ can be represented as expansions with coefficients dependent on $c_{LM}^E$ and $c_{LM}^B$. Therefore, the constraint on the Stokes parameters can be used to place constraints on the linear-polarization expansion coefficients $c_{LM}^{E,B}$. To place a simplified set of constraints, let us consider a background that has an isotropic intensity map, with deviation from isotropy sourced only by a single, nonzero, linear-polarization component $c_{L0}^{X}$ for $X \in \{E,\, B\}$. Then, the Stokes parameter constraint translates to:
\begin{equation}\label{eq: EB_constraint}
    c_{L0}^{X}\, {_{+ 4}Y_{LM}}(\hat{\Omega}) \leq \frac{1}{\sqrt{4\pi}}\,,
\end{equation}
where we have applied the assumed normalization of the intensity map ($c_{00}^I = 1$). Since the above relation must apply for any value of $\hat{\Omega}$, given the maximum of ${_{+ 4}Y_{LM}}(\hat{\Omega})$, one can derive the maximum allowed value $(c_{L0}^{E,B})_{\rm max}$. Note that, consistent with prior analysis \cite{Hotinli:2019tpc, Belgacem:2020nda}, we focus on the constraint for $M = 0$. A roughly similar bound applies to $\sqrt{2}\,{\rm Re}\,c_{LM}^X$ and $\sqrt{2}\,{\rm Im}\,c_{LM}^X$ for $M \neq 0$.

\section{Review of Bipolar Spherical Harmonics}
\label{sec:biposh}

We calculate our estimators in harmonic space, using the BiPoSH formalism described in previous works such as Refs.~\cite{Hotinli:2019tpc, Belgacem:2020nda}. Although this formalism relies on our ability to measure a well-distributed set of pulsars across the sky, it clearly depicts the contributions from the expansion coefficients of the Stokes parameters, revealing parameter degeneracies and simplified expressions for the relevant estimators. 

As emphasized in Ref.~\cite{Sato-Polito:2023spo}, the polarization anisotropies at different frequencies are not expected to be the same if the GW sources are SMBHBs.
Therefore, for our analysis, we work in a single frequency bin, dropping any explicit $f$ dependence, assuming we are dealing with maps of the timing residual $z(\hat{n})$ across the sky.

\label{subsec: biposh}
The spatial two-point correlation function of the timing residuals from two pulsars in directions $\hat n$ and $\hat m$ can (most generally) be expanded as,
\begin{eqnarray}\label{eq: biposh_exp_real}
 \VEV{z(\hat{n})z(\hat{m})} &= &\sum_{\ell}\frac{2\ell + 1}{4\pi}C_\ell P_\ell(\hat{n}\cdot\hat{m}) \\\nonumber
 &+ &\sum_{\ell\ell' LM} A^{LM}_{\ell\ell'}\left\{Y_{\ell}(\hat{n})\otimes Y_{\ell'}(\hat{m})\right\}_{LM}\,,
\end{eqnarray}
where $C_\ell$ and $A_{\ell\ell'}^{LM}$ are expansion coefficients and $P_{\ell}$ are Legendre polynomials. The second term in the equation above represents an expansion in the set of basis functions:
\begin{multline}
 \left\{Y_{\ell}(\hat{n})\otimes Y_{\ell'}(\hat{m})\right\}_{LM} = \\\sum_{mm'} \VEV{\ell m\ell' m'| LM}Y_{\ell m}(\hat{n})Y_{\ell' m'}(\hat{m})\,,
 \label{eq:biposh_def}
\end{multline}
called bipolar spherical harmonics \cite{Hajian:2003qq,Hajian:2005jh,Joshi:2009mj,Book:2011na}, where the symbol $\VEV{\ell m\ell' m'| LM}$ is used to represent Clebsch-Gordan coefficients. 
These BiPoSHs constitute a complete orthonormal basis for functions of $\hat{n}$ and $\hat{m}$ in terms of total-angular-momentum states of quantum numbers $LM$. 
The BiPoSH coefficients $A_{\ell\ell'}^{LM}$ are (anti)symmetric in $\ell$ and $\ell'$ for $\ell + \ell' + L =$ even (odd) when $z(\hat{n})$ is a real map \cite{Book:2011na}. These parity relations also indicate that $A_{\ell\ell'}^L$ must be zero for $\ell + \ell' + L =$ odd. Note that the sum in the second term in Eq.~\eqref{eq: biposh_exp_real} is over $L\geq 1$, $M=-L$ to $L$, and for values of $\ell,\ell'$ that satisfy the triangle inequality, $|\ell-\ell'| \leq L \leq \ell+\ell'$.

Given that the same angular correlation $\VEV{z(\hat{n})z(\hat{m})}$ can also be represented via
correlations of the spherical-harmonic coefficients $z_{\ell m}$ introduced in Eq.~\eqref{eq: single_resid_exp}, one can also express the two-point function in harmonic space as follows:
\begin{eqnarray}
 \VEV{z_{\ell m}^* z_{\ell 'm'}} & =& C_\ell \delta_{\ell \ell'} \delta_{mm'} \nonumber \\
 & + & \sum_{L\geq1}\sum_{M=-L}^L (-1)^{m} \VEV{\ell, -m, \ell'm'|LM} A^{LM}_{\ell\ell'}\,.\nonumber 
 \\
 \label{eq: biposh_exp_zlm}
\end{eqnarray}

The coefficients $C_\ell$, appearing in both Eq.~\eqref{eq: biposh_exp_real} and Eq.~\eqref{eq: biposh_exp_zlm}, represent the isotropic contribution to the background. That is, if we were analyzing an unpolarized, isotropic SGWB, then the only nonzero contribution to the angular correlation function would come from the power spectrum $C_\ell \propto (\ell -2)!/(\ell +2)!$, corresponding to the harmonic expansion of the Hellings-Downs function \cite{Gair:2014rwa,Roebber:2016jzl,Qin:2018yhy}.

On the other hand, the BiPoSH coefficients $A^{LM}_{\ell\ell'}$ quantify departures from statistical isotropy \cite{Hotinli:2019tpc}, circular-polarization anisotropies \cite{Belgacem:2020nda}, and (as we will show here) linear polarization \cite{Chu:2021krj,Liu:2022skj}. 
In other words, each term of given $LM$, in Eqs.~(\ref{eqn:expansion_I})--(\ref{eqn:expansion_QU}), gives rise to nonzero $A^{LM}_{\ell\ell'}$ of the same $LM$. 
As we will show in the following analysis, intensity anisotropies and $E$-mode linear polarization are scalars, and thus, have nonvanishing BiPoSH coefficients only for $\ell+\ell'+L=$ even, while $B$-mode linear polarization and circular polarization are pseudoscalars and thus are nonvanishing only for $\ell+\ell'+L=$odd. 
We will also see that the distinction between intensity anisotropies and $E$-mode polarization is found in the $\ell\ell'$ dependence of the BiPoSH coefficients, while the $B$-mode polarization and circular polarization are distinguished by whether the timing-residual maps are real or imaginary.

Under the assumption of a full-sky map of uniform noise properties, minimum-variance estimators for the BiPoSH coefficients have been previously computed and are given by 
\begin{eqnarray}
 \widehat{A^{LM}_{\ell\ell'}} = \sum_{mm'} z_{\ell m}^* z_{\ell 'm'}(-1)^{m}\VEV{\ell, -m,\ell' m'| LM}\,.
 \label{eq: A_LM_estimator}
\end{eqnarray}
Under the null hypothesis, corresponding to an isotropic, unpolarized background, this estimator has variance~\cite{Book:2011na},
\begin{equation}
 \VEV{\left|\widehat{A^{LM}_{\ell\ell'}}\right|^2} = \left[1+ (-1)^{\ell+\ell'+L}\delta_{\ell\ell'}\right]C_{\ell}C_{\ell'}.
 \label{eq: A_LM_estimator_variance}
\end{equation}
The above expression for the minimum variance has been simplified given that the covariance between BiPoSH coefficients $\VEV{A_{\ell\ell'}^{LM} (A_{\bar{\ell}\bar{\ell}'}^{LM})^*}$ is zero for $\{\ell, \ell'\} \neq \{\bar{\ell}, \bar{\ell}'\}$ \cite{Book:2011na}.  

In both the above equations, one can account for measurement noise by replacing $z_{\ell m}$ with the harmonic coefficients of the \textit{observed} timing residuals $z_{\ell m}^{\rm d} = z_{\ell m} + z_{\ell m}^{\rm noise}$. For the sake of our analysis, we assume a simplistic noise model in which pulsar red noise is omitted, and the coefficients $z_{\ell m}^{\rm noise}$ are uncorrelated, with variance given by the $\ell$-independent noise power spectrum $N_{zz}$. This white-noise power spectrum is a good approximation in the limit that the timing-residual noises in all pulsars are comparable. Consequently, to account for the effects of uncorrelated, Gaussian, white noise every occurrence of $C_{\ell}$ in estimator variances can be replaced by  $C_{\ell}^{\rm d}=C_\ell + N^{zz}$.

Note that the above expansion of the angular correlations in map $z(\hat{n})$ does not directly incorporate any specific model of the SGWB. 
Therefore, we need to rederive an expression for $\VEV{z_{\ell m}^* z_{\ell 'm'}}$ in terms of the relevant polarization expansion coefficients to leverage the estimator $\widehat{A^{LM}_{\ell\ell'}}$ and its previously computed variance in Eqs.~\eqref{eq: A_LM_estimator}~and~\eqref{eq: A_LM_estimator_variance}.

\section{Harmonic-space Analysis}
\label{sec:harmonicspace}

The next step is to determine the response of the signal $\VEV{z_{\ell m}^* z_{\ell 'm'}}$ to the intensity anisotropies and polarization, characterized by the expansion coefficients $c_{LM}^I$ and $c_{LM}^\pm$ [Eqs.~\eqref{eqn:expansion_I}~and~\eqref{eqn:expansion_QU}] in our model for the SGWB.

Following the derivations in Refs.~\cite{Hotinli:2019tpc, Belgacem:2020nda}, we begin by considering a GW propagating in the direction $\hat{\Omega} = \hat{z}$. Based on the imprint of a single GW on the pulse arrival time in configuration space [Eq.~\eqref{eq: timing_resid_oneGW}], we can derive the pulsar redshift response to a GW propagating in the $\hat{z}$ direction as:
\begin{equation}
 z(\hat{n}|\hat{z}) = \frac{1}{2}\left[\tilde{h}_{+}(1-\cos{\theta}) \cos{2\phi} + \tilde{h}_\times(1-\cos{\theta})\sin{2\phi}\right]\,,
\end{equation}
where $\tilde{h}_+(\hat{z})$ and $\tilde{h}_\times(\hat{z})$ are the Fourier amplitudes for the GW at a particular frequency $f$. 
Note that here, assuming that the wavelength of the GWs is significantly shorter than the Earth-pulsar distance ($fL \gg 1)$, we have dropped the subdominant, exponential `pulsar-term' in Eq.~\eqref{eq: single_residual_overlap_h}~\cite{2018JPhCo...2j5002M}.

The above expression can be used to derive the spherical harmonic coefficients
\begin{eqnarray}
 z_{\ell m}(\hat{z}) = \frac{z_{\ell}}{2}\left[(\td{h}_+ - i\td{h}_\times)\delta_{m2} + (\td{h}_+ + i\td{h}_\times)\delta_{m,-2}\right],
\end{eqnarray}
where $\delta_{ij}$ is the Kronecker delta and we have defined the symbol
\begin{eqnarray}
 z_{\ell} \equiv (-1)^\ell \sqrt{\frac{4\pi(2\ell + 1)(\ell - 2)!}{(\ell+2)!}}\,.
\end{eqnarray}
One can generalize the above result to a GW propagating in an arbitrary direction $\hat{\Omega}$ using Wigner rotation functions $D_{mm'}^{(\ell)}$ as follows \cite{Khersonskii:1988krb}:
\begin{eqnarray}
  z(\hat{n}| \hat{\Omega}) = \sum_{\ell = 2}^{\infty}\sum_{m m'}Y_{\ell m}(\hat{\Omega}) D_{mm'}^{(\ell)}(\hat{\Omega})\ z_{\ell m'}(\hat{z})\,,
\end{eqnarray}
where we have used $D_{mm'}^{(\ell)}(\hat{\Omega}) = D_{mm'}^{(\ell)}(\phi, \theta, 0)$ with $\hat{\Omega} = (\theta, \phi)$.\footnote{Note that even though a standard Wigner rotation involves three Euler angles, here we only rotate by $\theta$ and $\phi$ to ensure that the two polarizations `+' and `$\times$' remain aligned with $\hat{\theta}$ and $\hat{\phi}$. This Wigner function thus also rotates the dependence of the Fourier amplitudes $h_+$ and $h_\times$.} Therefore, based on Eq.~\eqref{eq: single_resid_exp}, we can see that the expansion coefficients for single GW traveling in direction $\hat{\Omega}$ can be expressed as:
\begin{eqnarray}
 z_{\ell m}(\hat{\Omega}) = \frac{z_\ell}{\sqrt{2}}\left[\tilde{h}_{L}(\hat{\Omega})D_{m2}^{(\ell)}(\hat{\Omega}) + \tilde{h}_{R}(\hat{\Omega})D_{m,-2}^{(\ell)}(\hat{\Omega})\right]\,, 
\end{eqnarray}
where we have defined $\tilde{h}_{R}(\hat{\Omega}) \equiv 1/\sqrt{2}(h_+ + ih_\times)(\hat{\Omega})$ and $\tilde{h}_{L}(\hat{\Omega}) \equiv 1/\sqrt{2}(h_+ - ih_\times)(\hat{\Omega})$ for ease of notation. Given this result, we can finally write the harmonic-space response of the PTA system to a background of GWs as
\begin{eqnarray}
 z_{\ell m} = \frac{z_{\ell}}{\sqrt{2}} \int \dd ^2\hat{\Omega}\left[\tilde{h}_{L}(\hat{\Omega})D_{m2}^{(\ell)} + \tilde{h}_{R}(\hat{\Omega})D_{m,-2}^{(\ell)}\right]\,.
\end{eqnarray}

Now, to calculate the correlation between two distinct harmonic coefficients, we plug in the statistical definition of the background from Eq.~\eqref{eq: GW_strain_correlations} and use the definition of the Stokes parameters [Eq.~\eqref{eqn:stokes}] to split the result into three separate contributions as follows:
\begin{eqnarray}
 \VEV{z_{\ell m}^* z_{\ell'm'}} = \frac{z_{\ell}z_{\ell'}}{2}\left(\mathcal{Z}_{\ell\ell'mm'}^{I} + \mathcal{Z}_{\ell\ell'mm'}^{P_+} +\mathcal{Z}_{\ell\ell'mm'}^{P_-}\right),
 \label{eq: harm_corr_components}
\end{eqnarray}
where we have assumed that background has no circular polarization. 
The first term corresponds to the contribution from the intensity of the background $I(\hat{\Omega})$ whereas the latter two terms represent the consolidated contributions from the linear polarization parameters $Q(\hat{\Omega})$ and $U(\hat{\Omega})$. Using Eqs.~\eqref{eqn:expansion_I} and~\eqref{eqn:expansion_QU}, one can write the separate contributions in terms of the expansion coefficients $c_{LM}^I$ and $c_{LM}^{\pm}$ as
\begin{align}
\label{eq:z_ll_mm_I}
 \mathcal{Z}_{\ell\ell'mm'}^{I} &= I(f) \sum_{LM}c_{LM}^I\int \dd ^{2}\hat{\Omega}\ Y_{LM}[(D_{m2}^{(\ell)})^*D_{m'2}^{(\ell')} \nonumber\\
 &\quad + (D_{m,-2}^{(\ell)})^*D_{m',-2}^{(\ell')}]\,, 
 \\
 \label{eq:z_ll_mm_QU}
 \mathcal{Z}_{\ell\ell'mm'}^{P_\pm} &= I(f) \sum_{LM}c_{LM}^\pm\int \dd^{2}\hat{\Omega}\ {_{\pm 4}}Y_{LM}(D^{(\ell)}_{m,\pm 2})^*D^{(\ell')}_{m',\mp 2}, 
\end{align}
where we have suppressed the $\hat{\Omega}$ dependence of the integrands for ease of notation. To calculate each independent contribution, we will need to compute the angular integrals over $\hat{\Omega}$. 

Given the analysis in Refs.~\cite{Hotinli:2019tpc, Belgacem:2020nda}, where they compute the term $\mathcal{Z}_{\ell\ell'mm'}^{I}$ above, we first focus on a background with an isotropic intensity map and compute only the $\mathcal{Z}_{\ell\ell'mm'}^{P_\pm}$ contributions. The required angular integrals can be computed by expressing the Wigner rotation functions in terms of spherical harmonics and then performing the integral over three spin-weighted spherical harmonics \cite{Gair:2015hra} which gives the following result:
\begin{eqnarray}
 \int \dd ^{2}\hat{\Omega}\ _{\pm 4}Y_{LM}(D^{(\ell)}_{m,\pm 2})^*D^{(\ell')}_{m',\mp 2} = \sqrt{4\pi}(-1)^{L + m} \\\nonumber \times \left( \begin{array}{ccc} \ell & \ell'& L \\ \pm 2 & \pm 2 & \mp 4 \end{array} \right) \VEV{\ell, -m, \ell'm'|LM}\,.
\end{eqnarray}
The above computation indicates that the two contributions $\mathcal{Z}_{\ell\ell' m m'}^{P_+}$ and $\mathcal{Z}_{\ell\ell' m m'}^{P_-}$ only differ in the lower line of the Wigner 3j symbol. Given the symmetry properties of these symbols, this is equivalent to a difference by a factor of $(-1)^{\ell + \ell' + L}$. 

Therefore, for a linearly polarized background characterized by Eq.~\eqref{eq: GW_strain_correlations}, the above calculations indicate that the anisotropic contribution from the Stokes parameters $Q$ and $U$ can be expanded in terms of Clebsch-Gordan coefficients, appearing precisely like the BiPoSH expansion introduced in Sec.~\ref{subsec: biposh}. Comparing the expansion in Eq.~\eqref{eq: biposh_exp_zlm} to the calculations above, we can relate the BiPoSH expansion coefficients $A_{\ell\ell'}^{LM}$ to $c_{LM}^{\pm}$ as follows:
\begin{eqnarray}
  A^{LM}_{\ell\ell'} = I(f) z_{\ell}z_{\ell'}\sqrt{4\pi}H_{\ell\ell'}^L\begin{cases} c_{LM}^E, & \ell+\ell'+L=\text{even}, \\
  i c_{LM}^B, & \ell+\ell'+L=\text{odd}, \end{cases} \nonumber\\
  \label{eq: biposh_All_clm_EB_rel}
\end{eqnarray}
where we have expressed $c_{LM}^{\pm}$ in terms of $c_{LM}^E$ and $c_{LM}^B$ using Eq.~\eqref{eq: plus_minus_to_EB} and we have defined
\begin{equation}
 H^{L}_{\ell\ell'} \equiv (-1)^L\left( \begin{array}{ccc} \ell & \ell'& L \\ 2 & 2 & -4 \end{array} \right)\,.
\end{equation}
Note that the lowest mode contributing to anisotropy in this setup is $L = 4$, since the linear combinations of $Q$ and $U$ are expanded in terms of spin-4 spherical harmonics.

\subsection{Fixing the normalization}
The sensitivity of a given PTA to anisotropy and/or polarization will depend on the number $N_p$ of pulsars and their noise properties. 
Strictly speaking, it will also depend on the location of the pulsars.
Here, we consider an optimized survey in which the pulsars are spread uniformly throughout the sky.\footnote{Real-world complications like irregular sky coverage and nonuniform noise will then only degrade the sensitivity relative to this optimal case.} 
Based on this setup, we follow Refs.~\cite{Hotinli:2019tpc,Belgacem:2020nda} in parametrizing the estimators of $c_{LM}^{E, B}$ in terms of the total SNR with which the {\it isotropic} signal is detected. Subsequently, this will allow us to write the normalization factor $I(f)$ in terms of the noise power spectrum and the isotropic SNR.

The measurement of the isotropic signal is contingent on the detection of the Hellings-Downs correlation, or equivalently, the power spectrum
\begin{eqnarray}
 C_{\ell} = \sqrt{4\pi}\frac{z_{\ell}^2}{(2\ell + 1)}I(f),
 \label{eq: Cl_If_rel}
\end{eqnarray}
where we have used Eqs.~\eqref{eq: harm_corr_components}~and~\eqref{eq:z_ll_mm_I} to obtain the normalization of $C_{\ell}$ for our model of the SGWB [Eq.~\eqref{eq: GW_strain_correlations}]. The total SNR for this isotropic measurement can be written as the sum in quadrature of the SNR of each accessible multipole $\ell \leq \ell_{\rm max}$, where $\ell_{\rm max} \sim \sqrt{N_p}$ is the largest accessible multipole. Assuming that the measurement noise is given by the white-noise power spectrum $N_{zz}$ (introduced below Eq.~\ref{eq: A_LM_estimator_variance}) this SNR can be written as:
\begin{eqnarray}
 {\rm SNR}_f = \left[\sum_{\ell = 2}^{\ell_{\rm max}}(2\ell + 1)\frac{C_{\ell}(f)^2}{(N^{zz})^2}\right]^{1/2} \approx \frac{(4\pi)^{3/2}}{6\sqrt{3}}\frac{I(f)}{N^{zz}}\,,
 \label{eq: SNR_If_rel}
\end{eqnarray}
where, the first equality has been previously derived in Ref.~\cite{Roebber:2016jzl}, and the second equality is has been derived numerically [using Eq.~\eqref{eq: Cl_If_rel}, as done in Ref.~\cite{Belgacem:2020nda}] given that the sum is dominated primarily by the lowest multipoles. This finally allows us to express the amplitude factor $I(f)$ as
\begin{eqnarray}
 I(f) = \frac{6\sqrt{3}}{(4\pi)^{3/2}} {\rm SNR}_fN_{zz}\,.
 \label{eq: SNR_Pf_rel}
\end{eqnarray}

\subsection{Harmonic Space Estimators}
\label{sec:harmonicspaceestimators}
Given the relation between the BiPoSH expansion coefficients $A^{LM}_{\ell\ell'}$ and the coefficients defining the statistics of the background $c_{LM}^{E,B}$ in Eq.~\eqref{eq: biposh_All_clm_EB_rel}, we can finally begin to calculate the estimators that quantify our ability to measure $Q$ and $U$ for a linearly polarized SGWB. 

The estimators for the expansion coefficients characterizing the Stokes parameters $Q$ and $U$ are most easy to calculate in the $E$- and $B$-mode basis. Each $\widehat{A^{LM}_{\ell\ell'}}$ provides one estimator:
\begin{eqnarray}
 \left(\widehat{c_{LM}^X}\right)_{\ell, \ell'} = \frac{1}{i^{X_{\ell\ell'}^L}\sqrt{4\pi} } \frac{\widehat{A^{LM}_{\ell\ell'}}}{H^{L}_{\ell\ell'} I(f) z_\ell z_{\ell'}}\,,
 \label{eq: est_cLM_EB_llp}
\end{eqnarray}
where $X = E$ when $\ell+\ell'+L$ is even and $X = B$ when $\ell+\ell'+L$ is odd. For concise notation, we have also defined $X_{\ell\ell'}^L \equiv (1/2)[1 + (-1)^{(\ell+\ell'+L)}]$. Given the variance of each BiPoSH amplitude [Eq.~\eqref{eq: A_LM_estimator_variance}], the variance in each of the above estimators is
\begin{eqnarray}
 \left(\Delta c_{LM}^X\right)_{\ell\ell'}^2 = \frac{4\pi^2}{27} \frac{\left[1 + (-1)^{X_{\ell\ell'}^L}\ \delta_{\ell\ell'}\right]C_{\ell}^{\rm d}C_{\ell'}^{\rm d}}{\left(z_{\ell}z_{\ell'} H_{\ell\ell'}^L {\rm SNR}_f N^{zz}\right)^2}\,,
 \label{eq: delta_cLM_EB_llp}
\end{eqnarray}
where we have plugged in Eq.~\eqref{eq: SNR_Pf_rel} and used $C_\ell^{\rm d} = C_{\ell} + N^{zz}$. 

Since each \textit{unique} pair $\ell, \ell'$ provides an estimator for the expansion coefficients\footnote{It is important to restrict the estimator sums to unique pairs of $\{\ell, \ell'\}$ to account for the (anti)symmetry of $A_{\ell,\ell'}^{LM}$ under $\ell \leftrightarrow \ell'$ for even (odd) parity BiPoSHs}, we can combine them with inverse variance weighting to construct the minimum variance estimator
\begin{eqnarray}
 \widehat{c_{LM}^X} = \frac{\sum_{\ell, \ell'}(\widehat{c_{LM}^X})_{\ell\ell'}\left(\Delta c_{LM}^X\right)_{\ell\ell'}^{-2}}{\sum_{\ell\ell'}\left(\Delta c_{LM}^X\right)_{\ell\ell'}^{-2}}\,,
 \label{eq: EB_min_var_est}
\end{eqnarray}
with the minimum variance given by
\begin{eqnarray}
 \left(\Delta c_{LM}^X\right)^{2} = \left[\sum_{\ell\ell'} \frac{27}{4\pi^2}\frac{\left(z_{\ell}z_{\ell'} H_{\ell\ell'}^L {\rm SNR}_f N^{zz}\right)^2}{\left[1 + (-1)^{X_{\ell\ell'}^L}\ \delta_{\ell\ell'}\right]C_{\ell}^{\rm d}C_{\ell'}^{\rm d}}\right]^{-1}\,.
 \label{eq: min_variance_fbin}
\end{eqnarray}
The above equation can therefore be used to obtain the estimator and corresponding variance of $c_{LM}^E$ ($c_{LM}^B$), by restricting the sums to \textit{unique} pairs of $\ell$ and $\ell'$ with even (odd) $\ell + \ell' + L$. Furthermore, in both the above equations, the sum is over $\ell, \ell' < \ell_{\rm max}$ such that $|\ell - \ell'| \leq L \leq \ell + \ell'$.

Finally, it turns out that the variance with which we can probe the expansion coefficients $c_{LM}^X$ can be written solely in terms of the SNR with which the isotropic contribution is measured. This can be achieved by using Eqs.~\eqref{eq: Cl_If_rel}~and~\eqref{eq: SNR_If_rel} to express
\begin{eqnarray}
 C_{\ell}^{\rm d} = C_{\ell} + N^{zz} = N^{zz}\left[\frac{6\sqrt{3}}{4\pi}\frac{z_\ell^2\ {\rm SNR}_f}{(2\ell + 1)} + 1\right]\,,
\end{eqnarray}
which eliminates the $N^{zz}$ dependence in Eq.~\eqref{eq: min_variance_fbin}. This also allows us to easily approximate how well we can measure the linear polarization expansion coefficients in the limit of low and high SNR. As the SNR$\to 0$, i.e., as the isotropic contribution of the background is measured with low fidelity, the variance with which one can estimate the expansion coefficients $c_{LM}^X$ is
\begin{eqnarray}
 \left(\Delta c_{LM}^X\right)^{2} \approx {\rm SNR}_f^{-2}\left\{\sum_{\ell\ell'} \frac{27}{4\pi^2}\frac{(z_\ell z_{\ell'}H_{\ell\ell'}^L)^2}{[1 + (-1)^{X_{\ell\ell'}^L} \delta_{\ell\ell'}]}\right\}^{-1}\,.
\end{eqnarray}
Conversely, as the SNR of the system becomes infinite, the error in the estimated expansion coefficient is
\begin{eqnarray}\label{eq: cLM_E_var_highSNR}
 \left(\Delta c_{LM}^X\right)^{2} \approx \left\{\sum_{\ell\ell'}\frac{(2\ell + 1)(2\ell' + 1)\left(H_{\ell\ell'}^L\right)^2}{[1 + (-1)^{X_{\ell\ell'}^L} \delta_{\ell\ell'}]}\right\}^{-1}\,.
\end{eqnarray}
Although the minimum variance calculated in the above analysis is purely real, it is important to remember that the coefficients $c_{LM}^X$ can have a nonzero imaginary part. Therefore, the minimum-variance estimator [Eq.~\eqref{eq: EB_min_var_est}] should be interpreted as an estimator for the real or imaginary part of the coefficients $c_{LM}^X$, each of which can be measured with minimum variance given by $1/2\left(\Delta c_{LM}^X\right)^{2}$. 

\subsection{Covariances}

The estimators calculated above show that the correlations induced by the $E$-mode polarization anisotropies differ from those of the $B$-mode component in the parity of the BiPoSHs allowed. Therefore, we should expect no covariance between the two components in a linearly polarized SGWB. 

However, these estimators were calculated assuming that the SGWB is isotropic in intensity and not circularly polarized. Most generally, the background may be anisotropic and partially polarized, with nonzero $I, Q, U$, and $V$ Stokes parameters. Intensity anisotropies are a scalar and give rise to nonzero $A^{LM}_{\ell\ell'}$ with $\ell+\ell'+L=$ even \cite{Hotinli:2019tpc}, while circular polarization is a pseudoscalar, giving rise to $A^{LM}_{\ell\ell'}$ with $\ell+\ell'+L=$ odd \cite{Belgacem:2020nda}. It is thus natural to expect some covariance between the intensity-anisotropy estimators and those for the linear-polarization $E$-mode, and we calculate this covariance below. We also show that the covariance between the circular-polarization estimators and the linear-polarization $B$-mode estimators vanishes, even though they have the same parity.

\subsubsection{Intensity and the $E$-Mode}

Let us consider a linearly polarized SGWB with an anisotropic intensity contribution. 
By evaluating the integrals in Eq.~\eqref{eq:z_ll_mm_I}, we can use the steps outlined for the previous estimator calculations to also write $\left(\widehat{c_{LM}^I}\right)_{\ell\ell'}$ in terms of the estimators of the BiPoSH coefficients as follows:
\begin{eqnarray}
 \left(\widehat{c_{LM}^I}\right)_{\ell\ell'} = \frac{1}{\sqrt{4\pi}z_\ell z_{\ell'}} \frac{\widehat{A^{LM}_{\ell\ell'}}}{G^{L}_{\ell\ell'} I(f)}\,,
 \label{eq: est_cLM_I_llp} 
\end{eqnarray}
for even $\ell + \ell' + L$, where we have defined the symbol
\begin{eqnarray}
 G^{L}_{\ell\ell'} \equiv (-1)^L\left( \begin{array}{ccc} \ell & \ell'& L \\ -2 & 2 & 0 \end{array} \right)\,.
\end{eqnarray}
Given the two estimator equations, presented in Eqs.~\eqref{eq: est_cLM_EB_llp}~and~\eqref{eq: est_cLM_I_llp}, the total covariance can be written as 
\begin{eqnarray}
 \VEV{\widehat{c_{LM}^E}\widehat{c_{LM}^I}} &=& \left[\sum_{\ell\ell'} \frac{27}{4\pi^2}\frac{\left(z_{\ell}z_{\ell'} {\rm SNR}_f N^{zz}\right)^2H_{\ell\ell'}^LG_{\ell\ell'}^L}{\left(1 + \delta_{\ell\ell'}\right)C_{\ell}^{\rm d}C_{\ell'}^{\rm d}}\right] \nonumber \\
 & & \times \left(\Delta c_{LM}^E \right)^{2} \left(\Delta c_{LM}^I \right)^{2}\,,
\label{eqn:covariance} 
\end{eqnarray}
where the sum is only over even values of $\ell +\ell' + L$ and we have used Eq.~(\ref{eq: SNR_Pf_rel}) to remove any frequency dependence beyond the SNR of the isotropic measurement in the specified frequency band.

We can then define a cross-correlation coefficient,
\begin{equation} \label{eq: r_def}
  r \equiv \frac{\VEV{\widehat{c_{LM}^E}\widehat{c_{LM}^I}}}{\left(\Delta c_{LM}^E \right) \left(\Delta c_{LM}^I \right)}\,.
\end{equation}
In the high-SNR limit, this becomes
\begin{eqnarray}
    r= \left[\sum_{\ell\ell'} \frac{(2\ell+1)(2\ell'+1)H^L_{\ell\ell'} G^L_{\ell\ell'}}{(1 + \delta_{\ell\ell'})}\right]\nonumber \\ \times \lim_{{\rm SNR}_f \rightarrow \infty} \left(\Delta c_{LM}^I \Delta c_{LM}^E\right)\,, \label{eqn:crosscorrelation} 
\end{eqnarray}
where $\lim_{\ {\rm SNR}_f \rightarrow \infty} \Delta c_{LM}^E$ can be obtained from Eq.~\eqref{eq: cLM_E_var_highSNR}, and one can use the same steps to calculate:
\begin{eqnarray}
    \lim_{{\rm SNR}_f \rightarrow \infty} \Delta c_{LM}^I = \left\{\sum_{\ell\ell'}\frac{(2\ell + 1)(2\ell' + 1)\left(G_{\ell\ell'}^L\right)^2}{(1 + \delta_{\ell\ell'})}\right\}^{-1/2}\,. \nonumber \\
\end{eqnarray}
Note that, throughout this subsection, we assume that all sums are restricted \textit{unique}, even-parity pairs $\{\ell, \ell'\}$ such that $\ell + \ell' + L = $ even. Furthermore, given a maximum multipole of HD-correlation detection $\ell_{\rm max}$ [Eq.~\eqref{eq: SNR_If_rel}], the sums are over $\ell, \ell' < \ell_{\rm max}$ such that $|\ell - \ell'| \leq L \leq \ell + \ell'$.

\subsubsection{Circular Polarization and the $B$-Mode}
To explore covariances across linear and circular polarization, let us consider a partially polarized SGWB with an isotropic intensity contribution and anisotropic linear- and circular-polarization maps. 
Given the SGWB defined in Sec.~\ref{sec: Characterizing the SGWB}, with nonzero $V(f, \hat{\Omega})$, one can write the estimator for $\left(\widehat{c_{LM}^V}\right)_{\ell, \ell'}$ in terms of the estimators of the BiPoSH coefficients as follows:
\begin{equation}
 \left(\widehat{c_{LM}^V}\right)_{\ell, \ell'} = \frac{1}{\sqrt{4\pi}z_\ell z_{\ell'}} \frac{\widehat{A^{LM}_{\ell\ell'}}}{G^{L}_{\ell\ell'} I(f)}\,,
\end{equation}
for $\ell + \ell' + L = $ odd. Although this estimator appears very similar to its intensity-map counterpart, there is a key difference in the properties of the BiPoSH coefficients appearing in the above equation.

An SGWB that has a circularly polarized component results in a timing residual map $z(\hat{n})$ that is, most generally, complex. This means that the harmonic space expansion coefficients $z_{\ell m}$ are not necessarily equal to $(-1)^mz_{\ell, -m}^*$. As a result, the odd-parity BiPoSH expansion coefficients $A_{\ell\ell'}^{LM}$ are no longer antisymmetric under $\ell \leftrightarrow \ell'$, and $A_{\ell\ell}^{LM}$ can be nonzero. Under these conditions, the odd-parity BiPoSH coefficients have variance given by~\cite{Belgacem:2020nda}
\begin{equation}
 \VEV{\left|\widehat{A^{LM}_{\ell\ell'}}\right|^2} = C_{\ell}C_{\ell'}\,.
\end{equation}
Therefore the variance in the estimator $\left(\widehat{c_{LM}^V}\right)_{\ell, \ell'}$ can be written as:
\begin{eqnarray}
 \left(\Delta c_{LM}^V\right)_{\ell\ell'}^2 = \frac{1}{4\pi}\frac{C_{\ell}C_{\ell'}}{\left[z_\ell z_{\ell'} G_{\ell\ell'}^L I(f)\right]^2},
\end{eqnarray}
which is symmetric in $\ell \leftrightarrow \ell'$. 

The complexity of the map, and the resulting loss of symmetry in the odd-parity BiPoSHs has a direct impact on the estimator for not only $c_{LM}^V$ but also $c_{LM}^B$, since, in each case, the estimator from the pair $\{\ell,\ell'\}$ is not degenerate with the estimator from $\{\ell', \ell\}$. Therefore, when constructing the estimators $\widehat{c_{LM}^V}$ and $\widehat{c_{LM}^B}$ for an imaginary timing residual map, we must include contributions from \textit{all} odd-parity $\ell, \ell'$ pairs with inverse variance weighting as follows:

\begin{eqnarray}
 \widehat{c_{LM}^V} = \left(\Delta c_{LM}^V\right)^{2}\sum_{\ell\ell'}\frac{\left(\widehat{A_{\ell\ell'}^{LM}} + \widehat{A_{\ell'\ell}^{LM}}\right)}{\sqrt{4\pi} z_\ell z_{\ell'}G_{\ell\ell'}^L I(f) \left(\Delta c_{LM}^V\right)^2_{\ell\ell'}}\,, \\
 \widehat{c_{LM}^B} = \left(\Delta c_{LM}^B\right)^{2}\sum_{\ell\ell'}\frac{\left(\widehat{A_{\ell\ell'}^{LM}} -\widehat{A_{\ell'\ell}^{LM}}\right) \times i^{X_{\ell\ell'}^L}}{\sqrt{4\pi} z_\ell z_{\ell'}H_{\ell\ell'}^L I(f) \left(\Delta c_{LM}^B\right)^2_{\ell\ell'}}\,,
\end{eqnarray}
where the sum is still over \textit{unique} sets of $\ell$ and $\ell'$ with $\ell + \ell' + L = $ odd. In each of the above equations, we have expanded the summation terms to account for the lack of symmetry in $A_{\ell\ell'}^{LM}$, using properties of the Wigner 3j symbols $H_{\ell\ell'}^L$ and $G_{\ell\ell'}^L$. The above estimators, therefore, indicate that the $B$-mode is sourced by the antisymmetric part of the odd-parity $A_{\ell\ell'}^{LM}$, whereas the circular polarization is sourced by the symmetric part of the same. Therefore, the covariance between $\widehat{c_{LM}^V}$ and $\widehat{c_{LM}^B}$ must be zero.  

\section{Configuration-Space Analysis: Overlap reduction functions}
\label{sec:configuration}

\begin{figure*}
 \centering
 \includegraphics[width=\columnwidth]{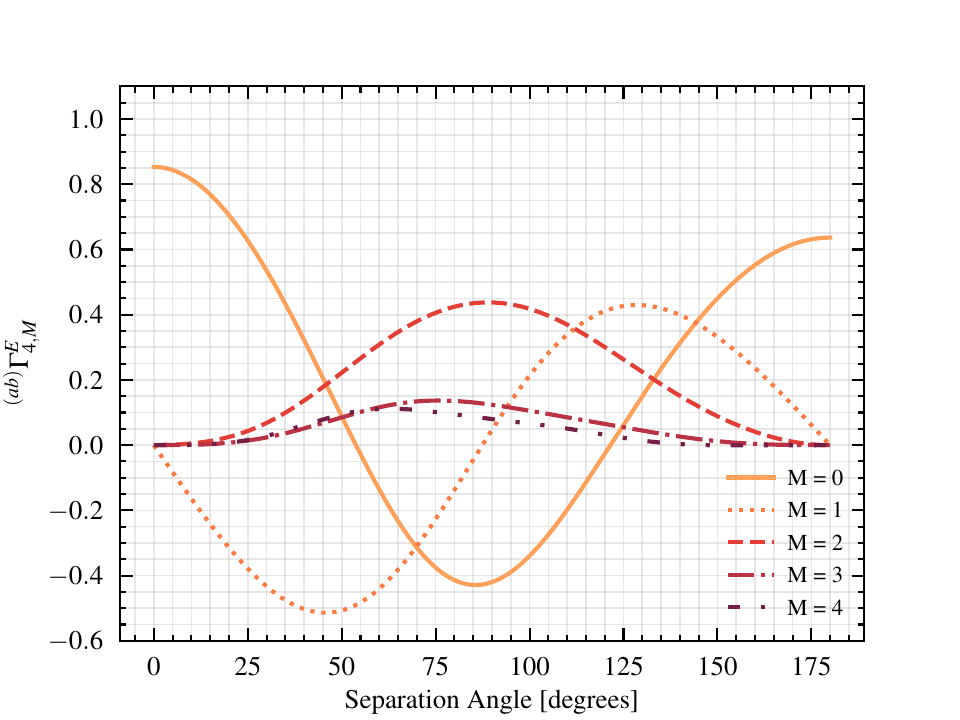}%
 \centering
 \includegraphics[width=\columnwidth]{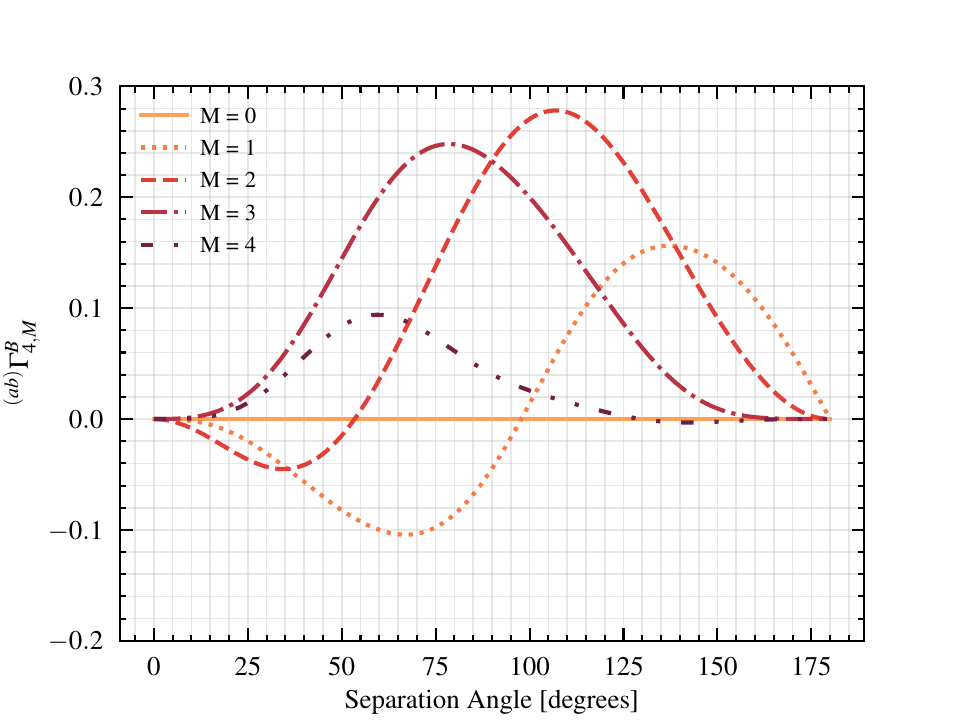}%
  \caption{\textit{Left:} overlap reduction functions for the $E$-mode polarization in the computational frame as a a function of pulsar separation angle for $L=4$. \textit{Right:} the same for the $B$-mode ORFs. The above results are calculated using the harmonic expansion presented in Eq.~\eqref{eqn:ORFs} Although the expression is coordinate independent, the results are plotted in the commonly used computational frame (defined in App.~\ref{App: ORF Computational}) to allow for easy comparison with previous, configuration-space evaluations of the same ORFs.}
 \label{fig:Gamma_4m_EB}%
\end{figure*}

The harmonic-space estimators derived above are uniquely useful because they allow for analytic estimates of the best sensitivity that a given PTA can achieve. However, realistic PTAs have irregular distributions of pulsars on the sky, with nonuniform noises. A configuration-space analysis allows one to account for these nonuniformities. In this regime, it is straightforward to derive constraints on linear-polarization anisotropies by repurposing pipelines that already exist to seek intensity anisotropy~(see, for example, Refs.~\cite{Mingarelli:2013dsa, Sato-Polito:2021efu}). The only additional step required is the computation of the relevant ORFs. 

Overlap reduction functions characterize the dependence of the two-point correlation function on the angular distance between the two chosen pulsars. The contribution from each of the four maps [expanded in Eqs.~(\ref{eqn:expansion_I} - \ref{eqn:expansion_QU})], to the total timing-residual correlation, is modulated by a unique ORF.
Although these functions can be computed in configuration space (see Appendix~\ref{App: ORF Computational}), in this section, we leverage the BiPoSH analysis above to derive the ORFs corresponding to the linear-polarization moments $c_{LM}^E$ and $c_{LM}^B$.

As above, suppose that the SGWB has an isotropic intensity contribution with an anisotropic linear polarization. 
The two-point correlation function can then be written as
\begin{align}
  \VEV{ z(\hat{n}_a) z(\hat{n}_b)} &= I(f)\zeta(\theta_{ab}) \\\nonumber
 &\quad + I(f) \sum_{LM}\left[c_{LM}^E{^{(ab)}}\Gamma^E_{L M} + ic_{LM}^B {^{(ab)}}\Gamma^B_{L M}\right],
\end{align}
where $\zeta(\theta_{ab})$ is the HD correlation for two sources separated by an angle $\theta_{ab}$; ${}^{(ab)}\Gamma_{LM}^E$ and ${}^{(ab)}\Gamma_{LM}^B$ are the overlap reduction functions for $E$- and $B$-mode linear polarization, respectively; and the sum on $L$ is over $L\geq4$.
Comparing the above expression with Eq.~(\ref{eq: biposh_exp_real}), we find that the ORFs can be expressed as
\begin{equation}
 {^{(ab)}}\Gamma^X_{L M} = \sum_{\ell\ell'} z_\ell z_{\ell'}\sqrt{4\pi}H_{\ell\ell'}^L\left\{Y_{\ell}(\hat{n}_a)\otimes Y_{\ell'}(\hat{n}_b)\right\}_{LM},
\label{eqn:ORFs} 
\end{equation}
where for $X=E$ ($X=B$) we sum over pairs $\ell,\ell'$ with $\ell + \ell' + L = $ even (odd). 
Here, we have to be careful about the summation terms, given that the ORF results are independent of the assumed polarization content of the SGWB. That is, these expansions hold true in the most general case of a polarized SGWB with contributions from all four Stokes parameters. As a result, to get appropriately normalized ORFs for the $E$- and $B$-mode expansion coefficients, we sum over $\textit{all}$ even and odd pairs  $\ell,\ell'$ (respectively) that satisfy the triangle inequality $|\ell - \ell'|\leq L\leq \ell + \ell'$.

This expression is far more economical and elegant than expressions for these ORFs in the prior literature. This form can be roughly explained as follows: If the anisotropy is decomposed into spherical harmonics of quantum numbers $LM$, rotational invariance (essentially, the Wigner-Eckart theorem) dictates that the two-point correlation function must take the form of something like Eq.~\eqref{eqn:ORFs}, as a sum over BiPoSHs with quantum numbers $LM\ell\ell'$. 
The physics is then entirely in the $L,\ell,\ell'$ dependence of the coefficient. Although Eq.~(\ref{eqn:ORFs}) is written formally as an infinite series, it is easily coded and numerically evaluated. 
For a survey with $N_p$ pulsars, the infinite series in $\ell,\ell'$ can be cut off at $\ell_{\rm max}\sim \sqrt{N_p}$. 
The simplicity of this result (and analogous results for intensity anisotropy and circular polarization) reduces the prospects of coding errors that may arise with complicated analytic formulas or inconsistencies between conventions used when taking results from different papers.

Since the above expansion is independent of any assumptions on the polarization content of the SGWB, the methodology presented here can be extended to intensity and circular polarization ORFs as well.
While this work shows a clearly derived example of the simplified ORF expressions for an anisotropic, linearly polarized background, Ref.~\cite{AnilKumar:2023yfw} shows how Eq.~(\ref{eqn:ORFs}) generalizes to intensity anisotropy and circular polarization and also provides the ORFs for a SGWB with the spin-1 GW polarizations that may arise in an alternative-gravity theory.

\section{Forecasts and Results}
\label{sec:forecasts}

\begin{figure*}
 \centering
 \includegraphics[width=\columnwidth]{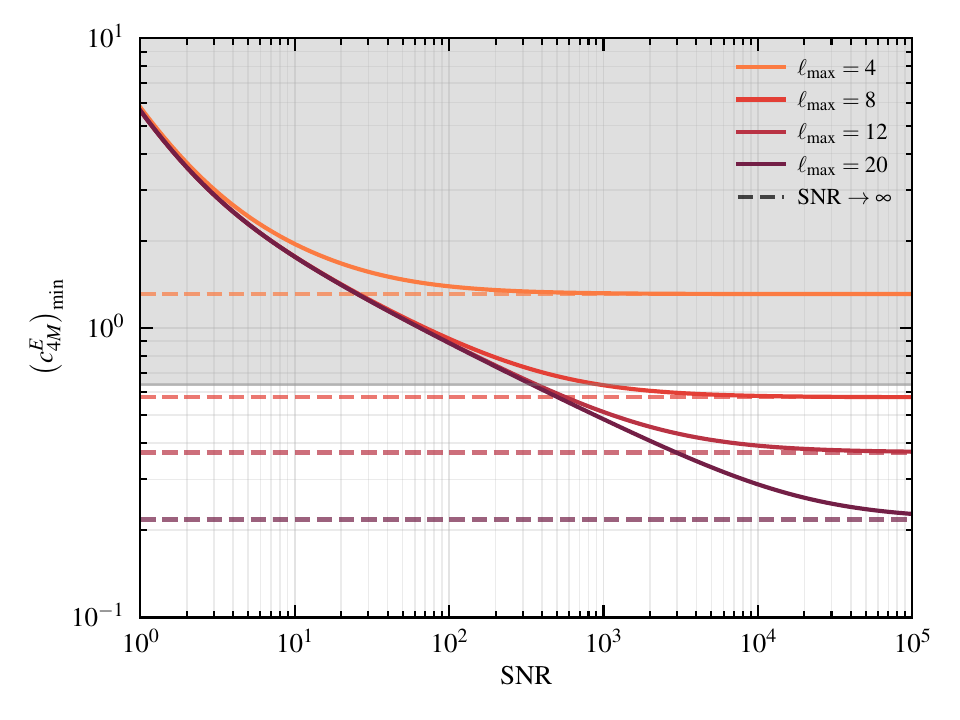}%
 \centering
 \includegraphics[width=\columnwidth]{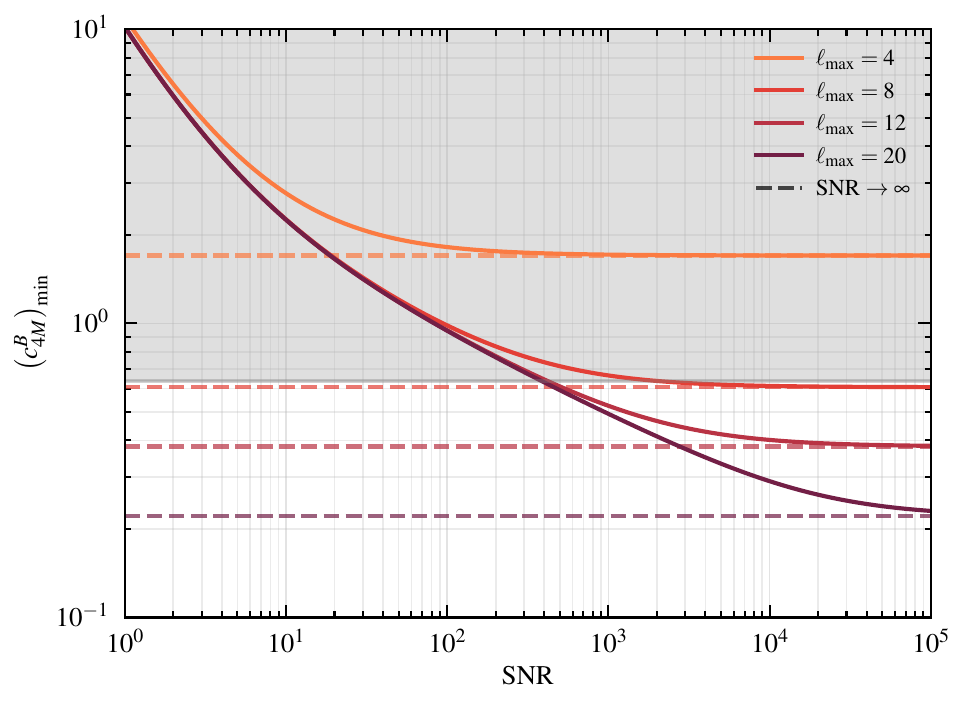}%
  \caption{\textit{Left:} smallest detectable (at 3-$\sigma$ level) $E$-mode amplitude $c_{4M}^E$ (real/ imaginary part) as a function of the SNR with which the Hellings-Downs curve is established. 
  Results are shown for different values of $\ell_{\rm max}\sim N_p^{1/2}$ (where $N_p$ is the number of pulsars). The dashed curves represent the high SNR limits, i.e., they show the smallest $E$-mode polarization anisotropy detectable by a PTA with a given number of pulsars at $L = 4$. \textit{Right:} the same results for $c_{4M}^B$. In both the plots, the shaded area in gray represents the region in which the minimum measurable polarization anisotropy $(c_{4M}^{E, B})_{\rm min}$ is greater than the maximum allowed value of this expansion coefficient $(c_{40}^{E,B})_{\rm max}$ [Eq.~\eqref{eq: EB_constraint}].}%
 \label{fig:c_4m_min_3sig}%
\end{figure*}

In this section, we use the estimators from Sec.~\ref{sec:harmonicspaceestimators} to forecast future measurement prospects of the coefficients $c_{LM}^E$ and $c_{LM}^B$ for a given PTA observation. We parametrize the experiment in terms of the maximum multipole moment $\ell_{\rm max}$ used in the analysis; this is expected to scale roughly as $\ell_{\rm max}\sim N_p^{1/2}$ with the number of pulsars $N_p$ in the survey. We parametrize the noise in terms of the SNR with which the isotropic HD correlation is detected in the given frequency bin.

We begin by presenting the ORFs for both the $E$- and $B$-mode in Fig.~\ref{fig:Gamma_4m_EB}. The ORFs plotted are for the lowest-order linear polarization anisotropy ($L = 4$) at various values of $M$. These results were obtained by computing Eq.~\eqref{eqn:ORFs} for $\ell_{\rm max} = 10$ in the computational frame, i.e., assuming that one of the pulsars is in the $\hat z$ direction and the other in the $x$-$z$ plane. Given the transformation from between the $E$- and $B$-mode basis to the $P^{\pm}$ basis [summarized in Eq.~\eqref{eq: plus_minus_to_EB}], our results agree with those previously calculated~\cite{Chu:2021krj,Liu:2022skj} directly in the configuration space. 

Our forecasts on future PTA sensitivity to linear polarization anistotropies are presented in Fig.~\ref{fig:c_4m_min_3sig}. Consistent with the displayed ORFs, we present measurement prospects of the coefficients $c_{LM}^{E, B}$ at $L=4$. The figure comprises two subplots, each demonstrating the minimum detectable amplitude of the real or imaginary components of the expansion coefficients $c_{LM}^{E,B}$ for a detection threshold of 3-sigma. 

The anticipated error margins for these measurements are calculated based on Eq.~\eqref{eq: min_variance_fbin}, under the null hypothesis, corresponding to an isotropic and unpolarized SGWB.
The dashed lines in the figure represent the high-SNR thresholds, indicating the smallest linear-polarization anisotropy detectable by a PTA with a given number of pulsars. The shaded area in gray represents the region in which the minimum detectable polarization anisotropy $(c_{4M}^{E,B})_{\rm min}$ is greater than the maximum allowed value $(c_{40}^{E, B})_{\rm max}$ calculated according to the constraint in Eq.~\eqref{eq: EB_constraint}. Our findings align with those for intensity anisotropy and circular polarization, indicating that the detection of linear polarization is challenging. Successful detection typically requires a high-SNR detection of the HD curve and a larger number of pulsars. Specifically, under the simplifying assumptions of our harmonic analysis, the measurement of an individual component $c_{4M}^{E, B}$ is only possible if the number of pulsars allows for an $\ell_{\rm max} > 8$ and the isotropic intensity contribution is measured with ${\rm SNR} \gtrsim 300$. 

\begin{figure}
    \centering
    \includegraphics[width=\columnwidth]{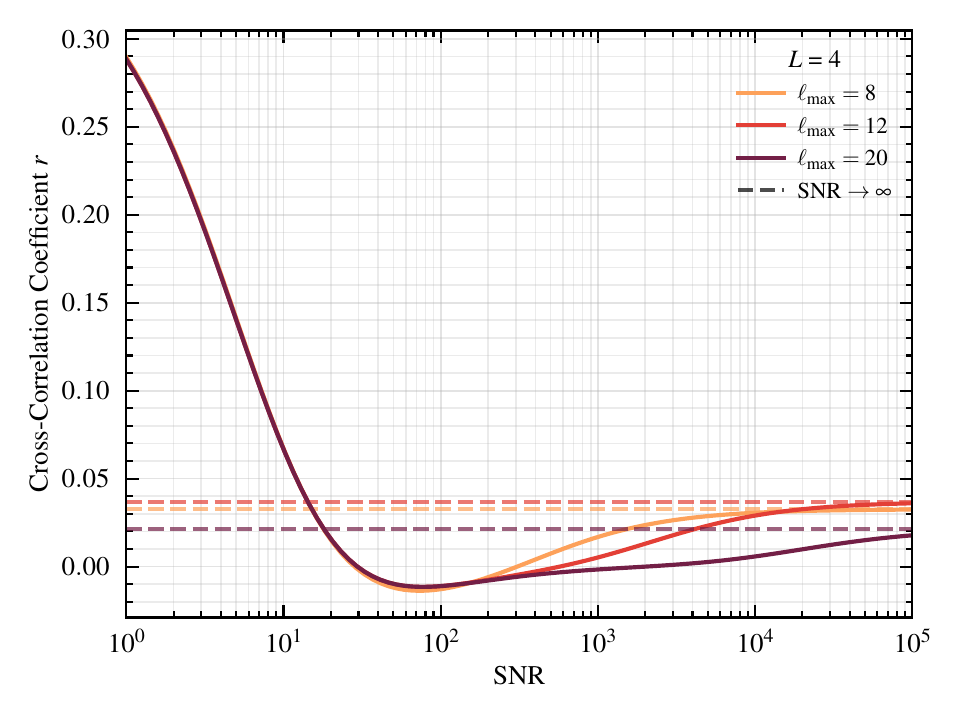}
    \caption{Cross-correlation coefficient $r$ [Eq.~\eqref{eq: r_def}] for $L=4$, quantifying the expected level of covariance between intensity anisotropy characterized by $c_{4M}^I$ and linear polarization $E$-mode anisotropy characterized by $c_{4M}^E$. Results are presented for various values of $\ell_{\rm max} \sim N_p^{1/2}$ (where $N_p$ is the number of pulsars) as a function of the isotropic-detection SNR (SNR of HD correlation detection). The dashed curves represent the high-SNR limits.}
    \label{fig: cross_corr_coef_L4}
\end{figure}

We also evaluate the cross-correlation coefficient $r$ [Eq.~\eqref{eq: r_def}] for $L=4$, which quantifies the expected covariance between the intensity anisotropy corresponding to $c_{4M}^I$ and the lowest order $E$-mode anisotropy characterized by $c_{4M}^E$. The results of this computation are presented in Fig.~\ref{fig: cross_corr_coef_L4}, where the solid lines represent the value of $r$ calculated using Eqs.~\eqref{eqn:covariance}~and~\eqref{eq: r_def} as a function of the isotropic-detection SNR for various values of $\ell_{\rm max}$. The results indicate that, when the HD correlation is measured with low SNR (${\rm SNR}_f \lesssim 100$), a larger number of pulsars does not alleviate the covariance between the two coefficients. Fortunately, this means that a PTA network can achieve a covariance below $5\%$ with a lower number of pulsars as long as $\ell_{\rm max} \geq 8$. However, the SNR of the isotropic detection must be at least 20. The dashed lines in the same plot indicate the value of $r$ under the high-SNR limit, calculated using Eq.~(\ref{eqn:crosscorrelation}). Although there is a separation in the results depending on the number of pulsars in the PTA, the expected covariance is small across all the assumed configurations, typically on the order of a few percent. Note that the required HD-detection SNR for the measurement of any individual linear-polarization anisotropy $c_{4M}^E$ smaller than $(c_{40}^E)_{\rm max}$ is higher than 100 [Fig.~\ref{fig:c_4m_min_3sig}]. Therefore, we conclude that with better measurements of the isotropic HD curve, the linear polarization and intensity anisotropy will be distinguishable.

\section{Conclusions}
\label{sec:conclusions}
Over the last decade, PTA collaborations have tirelessly worked toward the construction of a rigorous PTA network with increasingly precise timing residuals. Recently, these observations finally led to evidence for the existence of a stochastic signal, with correlations across pulsars following the expected Hellings-Downs curve \cite{1983ApJ...265L..39H}. These recent discoveries, alongside the prospects of expanding detection networks~\cite{Ng:2017djg, Bailes:2018azh} and uncovering new physics, call for a more detailed characterization of the SGWB.

Previous work has shown how intensity anisotropies and circular polarization can be sought with PTAs, both in terms of a formal harmonic-space description and in terms of the configuration-space ORFs used in data analysis. Here we have extended this earlier work to show how the linear polarization of the SGWB can be characterized. We provided economical derivations of results presented earlier \cite{Chu:2021krj,Liu:2022skj} and extended that work by providing estimators for the anisotropy amplitudes and the variances with which they can be measured. We use this formalism to forecast the minimum measurable amplitude of the linear polarization coefficients $c_{LM}^E$ and $c_{LM}^B$ (at $L=4$) as a function of the SNR with which the isotropic background is detected, for different values of $\ell_{\rm max}$. These results show that the linear polarization of the SGWB will only be accessible once the isotropic SNR is far larger, with the minimum detectable linear polarization decreasing with the square root of the number of pulsars.

In addition, we also show that there is a cross-correlation between the estimators for the intensity anisotropy and those for the $E$-mode linear polarization, but none between the circular-polarization estimators and those for the B-mode polarization, even though they have the same parity.
We find that the cross-correlation between the intensity anisotropy and linear polarization is small once the SNR becomes large, so it should be possible to separate the effects of the two polarization maps. Finally, we also provide a simple and elegant alternative for the computation of the linear-polarization ORFs.

Measurement of the linear polarization of the SGWB can offer additional information about the astrophysical sources of the background. It can be used as a tool to identify whether a small number of sources dominates the SGWB signal, possibly allowing for the identification of nearby individual binary systems. 
Since this contribution can naturally be present in PTA observations, characterizing the additional polarization components in the SGWB will become an important step in maximizing the useful information recovered from timing residual datasets.

\begin{acknowledgments}

We are grateful to Selim C. Hotinli for useful discussions. M.\c{C}.~is supported by NSF Grants No.~AST-2006538, PHY-2207502, PHY-090003 and PHY-20043; and by NASA Grants No.~20-LPS20-0011 and 21-ATP21-0010. This work was carried out at the Advanced Research Computing at Hopkins (ARCH) core facility (\url{rockfish.jhu.edu}), which is supported by the NSF Grant No.~OAC-1920103. This work was supported by NSF Grant No.~2112699, the Simons Foundation, and the John Templeton Foundation.

\end{acknowledgments}

\appendix
\section{Overlap Reduction Functions from Integrals}
\label{App: ORF Computational}
\begin{figure*}
 \centering
 \includegraphics[width=\columnwidth]{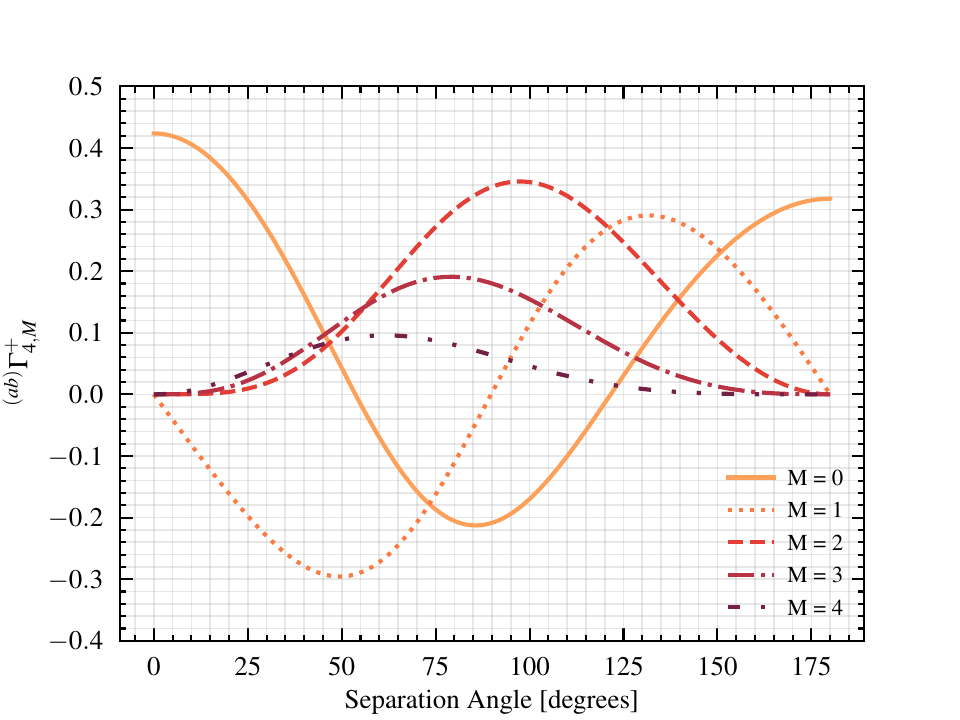}%
 \centering
 \includegraphics[width=\columnwidth]{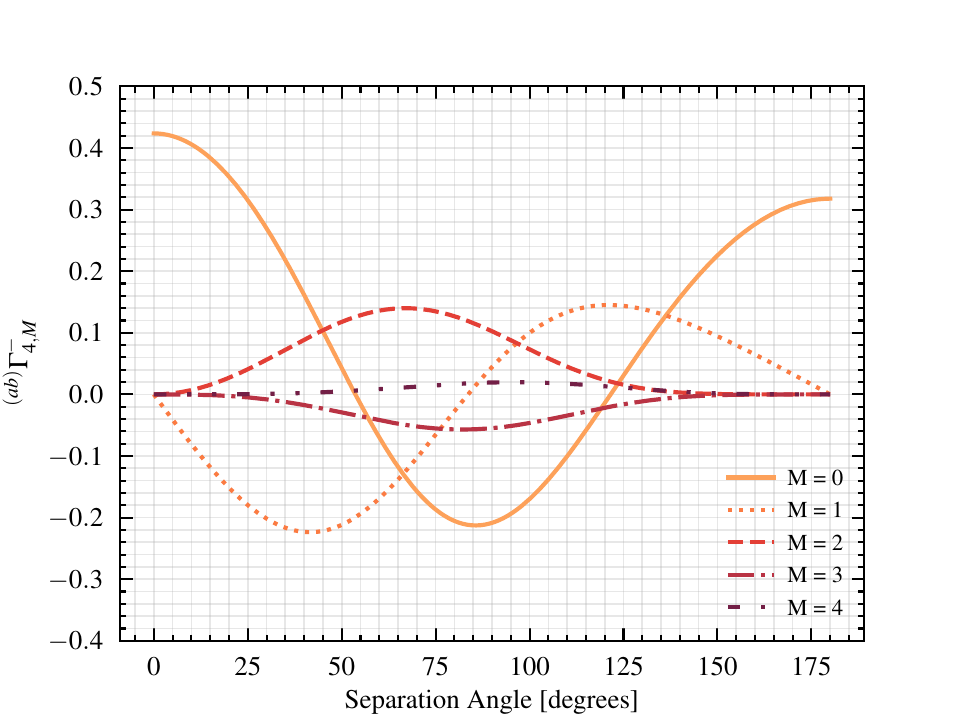}%
 \caption{\textit{Left:} overlap reduction functions for the $P_+$ polarization in the computational frame, as a function of the angle between the pulsars $a$ and $b$ for $L=4$. \textit{Right:} the same for the $P_-$ polarization. Both are calculated using numerical integration techniques applied to Eq.~\eqref{eq: ORF_integral}.}%
 \label{fig:Gamma_4m_pm_num}%
\end{figure*}

For completeness, we calculate here the ORFs in configuration space, as in previous work. It can be checked numerically that these expressions agree with the simpler representation in Eq.~(\ref{eqn:ORFs}).

For an SGWB with statistical properties defined in Eq.~\eqref{eq: GW_strain_correlations}, with the power-spectral-density tensor given in Eq.~\eqref{eq: polarization_corr_tensor}, the correlation induced across the timing residuals of two pulsars $a$ and $b$ is given by
\begin{align}\label{eq: tim_resid_long_SGWB}
\langle z(\hat{n}_a)z(\hat{n}_b) \rangle &= \int \dd ^2\hat{\Omega}\kappa_{ab}(\hat{\Omega}) 
\Big\{I(\hat{\Omega}) F_{ab}^I(\hat{\Omega}) \nonumber \\
&\quad + \frac{1}{2}P_+(\hat{\Omega}) F_{ab}^{P_+}(\hat{\Omega}) + \frac{1}{2}P_-(\hat{\Omega}) F_{ab}^{P_-}(\hat{\Omega})\Big\},
\end{align}
where we have suppressed the frequency dependence for ease of notation and we defined
\begin{eqnarray}
  F_{ab}^I(\hat{\Omega}) &= &F_a^+ F_b^+ + F_a^\times F_b^\times \nonumber\\
  F_{ab}^Q(\hat{\Omega}) &= &F_a^+ F_b^+ - F_a^\times F_b^\times \nonumber \\
  F_{ab}^U(\hat{\Omega}) &= &F_a^+ F_b^\times + F_a^\times F_b^+ \nonumber \\
  F_{ab}^{P_\pm}(\hat{\Omega}) & = &F_{ab}^Q(\hat{\Omega}) \mp i F_{ab}^U(\hat{\Omega}),
\end{eqnarray}
as well as
\begin{equation}
 \kappa_{ab}(f, \hat{\Omega}) \equiv \left[1-e^{-2\pi ifL_a(1 + \hat{\Omega}\cdot\hat{p}_a)}\right]\left[1-e^{2\pi ifL_b(1 + \hat{\Omega}\cdot\hat{p}_b)}\right].
\end{equation}

Note that the individual beam patterns $F^+(\hat{\Omega})$ and $F^\times(\hat{\Omega})$ are real for any chosen pulsar, and thus so are $F_{ab}^X(\hat{\Omega})$ ($X \in \{I,Q,U$\}). However, $F_{ab}^{P_\pm}(\hat{\Omega})$ are complex and are conjugates of each other. Consistent with prior analysis in the field, we work in the approximation that the period of the GWs in the nHz band is much smaller than the time taken for the signal to travel between Earth and the pulsar. This allows us to set $\kappa_{ab} \to (1+\delta_{ab})$.

As our next step, we plug in the angular expansions of the Stokes parameters presented in Eqs.~\eqref{eqn:expansion_I} and \eqref{eqn:expansion_QU}. This allows us to separate the frequency dependence of the background maps from the angular integral presented in Eq.~\eqref{eq: tim_resid_long_SGWB}. To simplify the resulting expression for the correlations, we characterize the ORFs:
\begin{eqnarray}
 {^{(ab)}}\Gamma^{X} = \sum_{LM} c_{LM}^X {^{(ab)}}\Gamma^{X}_{LM},
\end{eqnarray}
where $X \in \{I, P_+, P_-\}$ and we have defined
\begin{eqnarray}
 {^{(ab)}}\Gamma^I_{LM} = \int d^2\hat{\Omega}\ Y_{LM}(\hat{\Omega})\kappa_{ab} F_{ab}^I(\hat{\Omega}), \nonumber \\
 {^{(ab)}}\Gamma^\pm_{LM} = \frac{1}{2}\int d^2\hat{\Omega}\ {_{\pm 4}}Y_{LM}(\hat{\Omega})\kappa_{ab} F_{ab}^{P_{\pm}}(\hat{\Omega}).
 \label{eq: ORF_lm_elements}
\end{eqnarray}
At this point, it is worth explicitly pointing out that since $F_{ab}^{P_{+}} = (F_{ab}^{P_{-}})^*$ and ${_4}Y_{LM}^* = (-1)^m\,{_{-4}}Y_{L,-M}$, one can derive 
\begin{eqnarray}
 {^{(ab)}}\Gamma^-_{LM} = (-1)^m \left[{^{(ab)}}\Gamma^{+}_{L, -M}\right]^*\,.
 \label{eq: ORF_lin_relation}
\end{eqnarray}
Using the above ORFs, we can finally express the timing-residual correlation induced across pulsars $a$ and $b$ as:
\begin{widetext}
 \begin{eqnarray}
 \langle z_a^*(f)z_b(f')\rangle &= &\delta(f-f')\left[I(f)\sum_{L, M} c_{L M}^I {^{(ab)}}\Gamma^{I}_{LM}+I(f) \sum_{L, M} c_{LM}^+ {^{(ab)}}\Gamma^+_{LM} + I(f) \sum_{L, M} c_{LM}^- {^{(ab)}}\Gamma^-_{LM}\right] \\\nonumber
 &= &\delta(f-f')\left[I(f)\sum_{L, M} c_{LM}^I {^{(ab)}}\Gamma^{I}_{LM} + I(f) \sum_{L, M} c_{LM}^E {^{(ab)}}\Gamma^E_{LM}+ i I(f)\sum_{L, M} c_{LM}^B {^{(ab)}}\Gamma^B_{LM}\right].
\label{eqn:problematic}   
 \end{eqnarray}
\end{widetext}
To directly compare the above ORF calculations to their harmonic-space counterparts, in the second line of the equation above, we recast the correlation in terms of coefficients $c_{LM}^E$ and $c_{LM}^B$ using Eq.~\eqref{eq: plus_minus_to_EB}, with new ORF expansion functions,
\begin{eqnarray}
 {^{(ab)}}\Gamma^E_{LM} &= &{^{(ab)}}\Gamma^+_{LM} + {^{(ab)}}\Gamma^-_{LM}, \\\nonumber {^{(ab)}}\Gamma^B_{LM} &= &{^{(ab)}}\Gamma^+_{LM} - {^{(ab)}}\Gamma^-_{LM}.
 \label{eq: plus_minus_to_EB_ORFs}
\end{eqnarray}

To calculate the ORFs in configuration space, a coordinate system must be specified to define the pulsar positions and the gravitational wave propagation vector. This geometry defines the coordinate-dependent beam patterns [Eq.~\ref{eq: beam_pattern_def}] that contribute to the integrand of the ORF expansion functions. The calculations are most straightforward in the \textit{computational frame}, in which the $\hat{z}$-axis is aligned with one pulsar, with the other located on the $x-z$ plane, with an angular separation of $\zeta$. In this frame, the pulsar locations, GW propagation direction, and polarization are:
\begin{equation}
\begin{aligned}
 \hat{n}_a &= (0, 0, 1) \\
 \hat{n}_b &= \left(\sin{\zeta} , 0, \cos{\zeta}\right) \\
 \hat{\Omega} &= \left(\sin{\theta}\cos{\phi}, \sin{\theta}\sin{\phi}, \cos{\theta} \right) \\
 \hat{p} &= \left(\sin{\phi}, -\cos{\phi}, 0\right) \\
 \hat{q} &= \left(\cos{\theta}\cos{\phi}, \cos{\theta}\sin{\phi}, -\sin{\theta}\right).
\end{aligned}
\end{equation}
In this geometry, the individual pulsar beam pattern functions take the following form \cite{Sato-Polito:2021efu}:
\begin{equation}
\begin{aligned}
 F_a^+(\hat{\Omega}) &= -\frac{1}{2}\left(1 - \cos{\theta}\right) \\
 F^\times_a(\hat{\Omega}) &= 0 \\
 F_{b}^+(\hat{\Omega}) &= 
 \frac{(\sin{\phi}\sin{\zeta})^2 - (\sin{\zeta}\cos{\theta}\cos{\phi} - \sin{\theta}\cos{\zeta})^2}{2(1 + \cos{\theta}\cos{\zeta} + \sin{\theta}\sin{\zeta}\cos{\phi})} \\
 F_{b}^\times(\hat{\Omega}) &= 
 \frac{(\sin{\phi}\sin{\zeta})(\cos{\theta}\sin{\zeta}\cos{\phi} - \sin{\theta}\cos{\zeta})}{1 + \cos{\theta}\cos{\zeta} + \sin{\theta}\sin{\zeta}\cos{\phi}}.
\end{aligned}
\end{equation}
In the computational frame, the ORF expansion functions are completely real, and therefore the integral numerically computed to obtain ${^{(ab)}}\Gamma^\pm_{LM}(\zeta)$ is:

\begin{eqnarray}
  {^{(ab)}}\Gamma^\pm_{LM}(\zeta) = \int d^2\hat{\Omega} F_{ab}^Q\ {\rm Re}\left\{{_{\pm 4}}Y_{LM}\right\} \pm F_{ab}^U\ {\rm Im}\left\{{_{\pm 4}}Y_{LM}\right\}, \nonumber\\
  \label{eq: ORF_integral}
\end{eqnarray}
where the $\hat{\Omega}$ dependence of the integrand has been suppressed for ease of notation. The ORFs computed in the computational frame must then be rotated back into the cosmic-rest frame.

Reference~\cite{Chu:2021krj} computes these ORFs analytically in the computational frame. Their results match the plots presented in Fig.~\ref{fig:Gamma_4m_pm_num}, which display the linear polarization ORFs in the $P_+$ and $P_-$ basis computed in the computational frame using numerical integration techniques in Python. Given Eq.~\eqref{eq: plus_minus_to_EB_ORFs}, these results are consistent with the ORFs plotted in Fig.~\ref{fig:Gamma_4m_EB} using the harmonic expansion methodology from Eq.~\eqref{eqn:ORFs}.

\bibliography{PTA_linear}

\providecommand{\noopsort}[1]{}\providecommand{\singleletter}[1]{#1}%
\begin{thebibliography}{51}%
\makeatletter
\providecommand \@ifxundefined [1]{%
 \@ifx{#1\undefined}
}%
\providecommand \@ifnum [1]{%
 \ifnum #1\expandafter \@firstoftwo
 \else \expandafter \@secondoftwo
 \fi
}%
\providecommand \@ifx [1]{%
 \ifx #1\expandafter \@firstoftwo
 \else \expandafter \@secondoftwo
 \fi
}%
\providecommand \natexlab [1]{#1}%
\providecommand \enquote  [1]{``#1''}%
\providecommand \bibnamefont  [1]{#1}%
\providecommand \bibfnamefont [1]{#1}%
\providecommand \citenamefont [1]{#1}%
\providecommand \href@noop [0]{\@secondoftwo}%
\providecommand \href [0]{\begingroup \@sanitize@url \@href}%
\providecommand \@href[1]{\@@startlink{#1}\@@href}%
\providecommand \@@href[1]{\endgroup#1\@@endlink}%
\providecommand \@sanitize@url [0]{\catcode `\\12\catcode `\$12\catcode
  `\&12\catcode `\#12\catcode `\^12\catcode `\_12\catcode `\%12\relax}%
\providecommand \@@startlink[1]{}%
\providecommand \@@endlink[0]{}%
\providecommand \url  [0]{\begingroup\@sanitize@url \@url }%
\providecommand \@url [1]{\endgroup\@href {#1}{\urlprefix }}%
\providecommand \urlprefix  [0]{URL }%
\providecommand \Eprint [0]{\href }%
\providecommand \doibase [0]{https://doi.org/}%
\providecommand \selectlanguage [0]{\@gobble}%
\providecommand \bibinfo  [0]{\@secondoftwo}%
\providecommand \bibfield  [0]{\@secondoftwo}%
\providecommand \translation [1]{[#1]}%
\providecommand \BibitemOpen [0]{}%
\providecommand \bibitemStop [0]{}%
\providecommand \bibitemNoStop [0]{.\EOS\space}%
\providecommand \EOS [0]{\spacefactor3000\relax}%
\providecommand \BibitemShut  [1]{\csname bibitem#1\endcsname}%
\let\auto@bib@innerbib\@empty
\bibitem [{\citenamefont {Agazie}\ \emph
  {et~al.}(2023{\natexlab{a}})\citenamefont {Agazie} \emph
  {et~al.}}]{NanoGrav:2023gor}%
  \BibitemOpen
  \bibfield  {author} {\bibinfo {author} {\bibfnamefont {G.}~\bibnamefont
  {Agazie}} \emph {et~al.} (\bibinfo {collaboration} {NANOGrav}),\ }\bibfield
  {title} {\bibinfo {title} {{The NANOGrav 15 yr Data Set: Evidence for a
  Gravitational-wave Background}},\ }\href
  {https://doi.org/10.3847/2041-8213/acdac6} {\bibfield  {journal} {\bibinfo
  {journal} {Astrophys. J. Lett.}\ }\textbf {\bibinfo {volume} {951}},\
  \bibinfo {pages} {L8} (\bibinfo {year} {2023}{\natexlab{a}})},\ \Eprint
  {https://arxiv.org/abs/2306.16213} {arXiv:2306.16213 [astro-ph.HE]}
  \BibitemShut {NoStop}%
\bibitem [{\citenamefont {Antoniadis}\ \emph {et~al.}(2023)\citenamefont
  {Antoniadis} \emph {et~al.}}]{EPTA:2023fyk}%
  \BibitemOpen
  \bibfield  {author} {\bibinfo {author} {\bibfnamefont {J.}~\bibnamefont
  {Antoniadis}} \emph {et~al.} (\bibinfo {collaboration} {EPTA}),\ }\bibfield
  {title} {\bibinfo {title} {{The second data release from the European Pulsar
  Timing Array III. Search for gravitational wave signals}},\ }\href
  {https://doi.org/10.1051/0004-6361/202346844} {\bibfield  {journal} {\bibinfo
   {journal} {Astron. Astrophys.}\ }\textbf {\bibinfo {volume} {678}},\
  \bibinfo {pages} {A50} (\bibinfo {year} {2023})},\ \Eprint
  {https://arxiv.org/abs/2306.16214} {arXiv:2306.16214 [astro-ph.HE]}
  \BibitemShut {NoStop}%
\bibitem [{\citenamefont {Reardon}\ \emph {et~al.}(2023)\citenamefont {Reardon}
  \emph {et~al.}}]{Reardon:2023gzh}%
  \BibitemOpen
  \bibfield  {author} {\bibinfo {author} {\bibfnamefont {D.~J.}\ \bibnamefont
  {Reardon}} \emph {et~al.},\ }\bibfield  {title} {\bibinfo {title} {{Search
  for an Isotropic Gravitational-wave Background with the Parkes Pulsar Timing
  Array}},\ }\href {https://doi.org/10.3847/2041-8213/acdd02} {\bibfield
  {journal} {\bibinfo  {journal} {Astrophys. J. Lett.}\ }\textbf {\bibinfo
  {volume} {951}},\ \bibinfo {pages} {L6} (\bibinfo {year} {2023})},\ \Eprint
  {https://arxiv.org/abs/2306.16215} {arXiv:2306.16215 [astro-ph.HE]}
  \BibitemShut {NoStop}%
\bibitem [{\citenamefont {Xu}\ \emph {et~al.}(2023)\citenamefont {Xu} \emph
  {et~al.}}]{Xu:2023wog}%
  \BibitemOpen
  \bibfield  {author} {\bibinfo {author} {\bibfnamefont {H.}~\bibnamefont {Xu}}
  \emph {et~al.},\ }\bibfield  {title} {\bibinfo {title} {{Searching for the
  Nano-Hertz Stochastic Gravitational Wave Background with the Chinese Pulsar
  Timing Array Data Release I}},\ }\href
  {https://doi.org/10.1088/1674-4527/acdfa5} {\bibfield  {journal} {\bibinfo
  {journal} {Res. Astron. Astrophys.}\ }\textbf {\bibinfo {volume} {23}},\
  \bibinfo {pages} {075024} (\bibinfo {year} {2023})},\ \Eprint
  {https://arxiv.org/abs/2306.16216} {arXiv:2306.16216 [astro-ph.HE]}
  \BibitemShut {NoStop}%
\bibitem [{\citenamefont {Rajagopal}\ and\ \citenamefont
  {Romani}(1995)}]{Rajagopal:1994zj}%
  \BibitemOpen
  \bibfield  {author} {\bibinfo {author} {\bibfnamefont {M.}~\bibnamefont
  {Rajagopal}}\ and\ \bibinfo {author} {\bibfnamefont {R.~W.}\ \bibnamefont
  {Romani}},\ }\bibfield  {title} {\bibinfo {title} {{Ultralow frequency
  gravitational radiation from massive black hole binaries}},\ }\href
  {https://doi.org/10.1086/175813} {\bibfield  {journal} {\bibinfo  {journal}
  {Astrophys. J.}\ }\textbf {\bibinfo {volume} {446}},\ \bibinfo {pages} {543}
  (\bibinfo {year} {1995})},\ \Eprint {https://arxiv.org/abs/astro-ph/9412038}
  {arXiv:astro-ph/9412038} \BibitemShut {NoStop}%
\bibitem [{\citenamefont {Jaffe}\ and\ \citenamefont
  {Backer}(2003)}]{Jaffe:2002rt}%
  \BibitemOpen
  \bibfield  {author} {\bibinfo {author} {\bibfnamefont {A.~H.}\ \bibnamefont
  {Jaffe}}\ and\ \bibinfo {author} {\bibfnamefont {D.~C.}\ \bibnamefont
  {Backer}},\ }\bibfield  {title} {\bibinfo {title} {{Gravitational waves probe
  the coalescence rate of massive black hole binaries}},\ }\href
  {https://doi.org/10.1086/345443} {\bibfield  {journal} {\bibinfo  {journal}
  {Astrophys. J.}\ }\textbf {\bibinfo {volume} {583}},\ \bibinfo {pages} {616}
  (\bibinfo {year} {2003})},\ \Eprint {https://arxiv.org/abs/astro-ph/0210148}
  {arXiv:astro-ph/0210148} \BibitemShut {NoStop}%
\bibitem [{\citenamefont {{Hellings}}\ and\ \citenamefont
  {{Downs}}(1983)}]{1983ApJ...265L..39H}%
  \BibitemOpen
  \bibfield  {author} {\bibinfo {author} {\bibfnamefont {R.~W.}\ \bibnamefont
  {{Hellings}}}\ and\ \bibinfo {author} {\bibfnamefont {G.~S.}\ \bibnamefont
  {{Downs}}},\ }\bibfield  {title} {\bibinfo {title} {{Upper limits on the
  isotropic gravitational radiation background from pulsar timing analysis.}},\
  }\href {https://doi.org/10.1086/183954} {\bibfield  {journal} {\bibinfo
  {journal} {\apjl}\ }\textbf {\bibinfo {volume} {265}},\ \bibinfo {pages}
  {L39} (\bibinfo {year} {1983})}\BibitemShut {NoStop}%
\bibitem [{\citenamefont {Arzoumanian}\ \emph {et~al.}(2020)\citenamefont
  {Arzoumanian} \emph {et~al.}}]{NANOGrav:2020bcs}%
  \BibitemOpen
  \bibfield  {author} {\bibinfo {author} {\bibfnamefont {Z.}~\bibnamefont
  {Arzoumanian}} \emph {et~al.} (\bibinfo {collaboration} {NANOGrav}),\
  }\bibfield  {title} {\bibinfo {title} {{The NANOGrav 12.5 yr Data Set: Search
  for an Isotropic Stochastic Gravitational-wave Background}},\ }\href
  {https://doi.org/10.3847/2041-8213/abd401} {\bibfield  {journal} {\bibinfo
  {journal} {Astrophys. J. Lett.}\ }\textbf {\bibinfo {volume} {905}},\
  \bibinfo {pages} {L34} (\bibinfo {year} {2020})},\ \Eprint
  {https://arxiv.org/abs/2009.04496} {arXiv:2009.04496 [astro-ph.HE]}
  \BibitemShut {NoStop}%
\bibitem [{\citenamefont {Chen}\ \emph {et~al.}(2021)\citenamefont {Chen} \emph
  {et~al.}}]{Chen:2021rqp}%
  \BibitemOpen
  \bibfield  {author} {\bibinfo {author} {\bibfnamefont {S.}~\bibnamefont
  {Chen}} \emph {et~al.},\ }\bibfield  {title} {\bibinfo {title}
  {{Common-red-signal analysis with 24-yr high-precision timing of the European
  Pulsar Timing Array: inferences in the stochastic gravitational-wave
  background search}},\ }\href {https://doi.org/10.1093/mnras/stab2833}
  {\bibfield  {journal} {\bibinfo  {journal} {Mon. Not. Roy. Astron. Soc.}\
  }\textbf {\bibinfo {volume} {508}},\ \bibinfo {pages} {4970} (\bibinfo {year}
  {2021})},\ \Eprint {https://arxiv.org/abs/2110.13184} {arXiv:2110.13184
  [astro-ph.HE]} \BibitemShut {NoStop}%
\bibitem [{\citenamefont {Goncharov}\ \emph {et~al.}(2021)\citenamefont
  {Goncharov} \emph {et~al.}}]{Goncharov:2021oub}%
  \BibitemOpen
  \bibfield  {author} {\bibinfo {author} {\bibfnamefont {B.}~\bibnamefont
  {Goncharov}} \emph {et~al.},\ }\bibfield  {title} {\bibinfo {title} {{On the
  Evidence for a Common-spectrum Process in the Search for the Nanohertz
  Gravitational-wave Background with the Parkes Pulsar Timing Array}},\ }\href
  {https://doi.org/10.3847/2041-8213/ac17f4} {\bibfield  {journal} {\bibinfo
  {journal} {Astrophys. J. Lett.}\ }\textbf {\bibinfo {volume} {917}},\
  \bibinfo {pages} {L19} (\bibinfo {year} {2021})},\ \Eprint
  {https://arxiv.org/abs/2107.12112} {arXiv:2107.12112 [astro-ph.HE]}
  \BibitemShut {NoStop}%
\bibitem [{\citenamefont {Antoniadis}\ \emph {et~al.}(2022)\citenamefont
  {Antoniadis} \emph {et~al.}}]{Antoniadis:2022pcn}%
  \BibitemOpen
  \bibfield  {author} {\bibinfo {author} {\bibfnamefont {J.}~\bibnamefont
  {Antoniadis}} \emph {et~al.},\ }\bibfield  {title} {\bibinfo {title} {{The
  International Pulsar Timing Array second data release: Search for an
  isotropic gravitational wave background}},\ }\href
  {https://doi.org/10.1093/mnras/stab3418} {\bibfield  {journal} {\bibinfo
  {journal} {Mon. Not. Roy. Astron. Soc.}\ }\textbf {\bibinfo {volume} {510}},\
  \bibinfo {pages} {4873} (\bibinfo {year} {2022})},\ \Eprint
  {https://arxiv.org/abs/2201.03980} {arXiv:2201.03980 [astro-ph.HE]}
  \BibitemShut {NoStop}%
\bibitem [{\citenamefont {Bernardo}\ and\ \citenamefont
  {Ng}(2023{\natexlab{a}})}]{Bernardo:2023mxc}%
  \BibitemOpen
  \bibfield  {author} {\bibinfo {author} {\bibfnamefont {R.~C.}\ \bibnamefont
  {Bernardo}}\ and\ \bibinfo {author} {\bibfnamefont {K.-W.}\ \bibnamefont
  {Ng}},\ }\bibfield  {title} {\bibinfo {title} {{Constraining gravitational
  wave propagation using pulsar timing array correlations}},\ }\href
  {https://doi.org/10.1103/PhysRevD.107.L101502} {\bibfield  {journal}
  {\bibinfo  {journal} {Phys. Rev. D}\ }\textbf {\bibinfo {volume} {107}},\
  \bibinfo {pages} {L101502} (\bibinfo {year} {2023}{\natexlab{a}})},\ \Eprint
  {https://arxiv.org/abs/2302.11796} {arXiv:2302.11796 [gr-qc]} \BibitemShut
  {NoStop}%
\bibitem [{\citenamefont {Bernardo}\ and\ \citenamefont
  {Ng}(2023{\natexlab{b}})}]{Bernardo:2023bqx}%
  \BibitemOpen
  \bibfield  {author} {\bibinfo {author} {\bibfnamefont {R.~C.}\ \bibnamefont
  {Bernardo}}\ and\ \bibinfo {author} {\bibfnamefont {K.-W.}\ \bibnamefont
  {Ng}},\ }\bibfield  {title} {\bibinfo {title} {{Hunting the stochastic
  gravitational wave background in pulsar timing array cross correlations
  through theoretical uncertainty}},\ }\href
  {https://doi.org/10.1088/1475-7516/2023/08/028} {\bibfield  {journal}
  {\bibinfo  {journal} {JCAP}\ }\textbf {\bibinfo {volume} {08}},\ \bibinfo
  {pages} {028}},\ \Eprint {https://arxiv.org/abs/2304.07040} {arXiv:2304.07040
  [gr-qc]} \BibitemShut {NoStop}%
\bibitem [{\citenamefont {Bernardo}\ and\ \citenamefont
  {Ng}(2023{\natexlab{c}})}]{Bernardo:2023pwt}%
  \BibitemOpen
  \bibfield  {author} {\bibinfo {author} {\bibfnamefont {R.~C.}\ \bibnamefont
  {Bernardo}}\ and\ \bibinfo {author} {\bibfnamefont {K.-W.}\ \bibnamefont
  {Ng}},\ }\bibfield  {title} {\bibinfo {title} {{Testing gravity with cosmic
  variance-limited pulsar timing array correlations}},\ }\href@noop {} {\
  (\bibinfo {year} {2023}{\natexlab{c}})},\ \Eprint
  {https://arxiv.org/abs/2306.13593} {arXiv:2306.13593 [gr-qc]} \BibitemShut
  {NoStop}%
\bibitem [{\citenamefont {Bernardo}\ and\ \citenamefont
  {Ng}(2022)}]{Bernardo:2022xzl}%
  \BibitemOpen
  \bibfield  {author} {\bibinfo {author} {\bibfnamefont {R.~C.}\ \bibnamefont
  {Bernardo}}\ and\ \bibinfo {author} {\bibfnamefont {K.-W.}\ \bibnamefont
  {Ng}},\ }\bibfield  {title} {\bibinfo {title} {{Pulsar and cosmic variances
  of pulsar timing-array correlation measurements of the stochastic
  gravitational wave background}},\ }\href
  {https://doi.org/10.1088/1475-7516/2022/11/046} {\bibfield  {journal}
  {\bibinfo  {journal} {JCAP}\ }\textbf {\bibinfo {volume} {11}},\ \bibinfo
  {pages} {046}},\ \Eprint {https://arxiv.org/abs/2209.14834} {arXiv:2209.14834
  [gr-qc]} \BibitemShut {NoStop}%
\bibitem [{\citenamefont {Agazie}\ \emph {et~al.}(2024)\citenamefont {Agazie}
  \emph {et~al.}}]{NANOGrav:2023ygs}%
  \BibitemOpen
  \bibfield  {author} {\bibinfo {author} {\bibfnamefont {G.}~\bibnamefont
  {Agazie}} \emph {et~al.} (\bibinfo {collaboration} {NANOGrav}),\ }\bibfield
  {title} {\bibinfo {title} {{The NANOGrav 15 yr Data Set: Search for
  Transverse Polarization Modes in the Gravitational-wave Background}},\ }\href
  {https://doi.org/10.3847/2041-8213/ad2a51} {\bibfield  {journal} {\bibinfo
  {journal} {Astrophys. J. Lett.}\ }\textbf {\bibinfo {volume} {964}},\
  \bibinfo {pages} {L14} (\bibinfo {year} {2024})},\ \Eprint
  {https://arxiv.org/abs/2310.12138} {arXiv:2310.12138 [gr-qc]} \BibitemShut
  {NoStop}%
\bibitem [{\citenamefont {Qin}\ \emph {et~al.}(2019)\citenamefont {Qin},
  \citenamefont {Boddy}, \citenamefont {Kamionkowski},\ and\ \citenamefont
  {Dai}}]{Qin:2018yhy}%
  \BibitemOpen
  \bibfield  {author} {\bibinfo {author} {\bibfnamefont {W.}~\bibnamefont
  {Qin}}, \bibinfo {author} {\bibfnamefont {K.~K.}\ \bibnamefont {Boddy}},
  \bibinfo {author} {\bibfnamefont {M.}~\bibnamefont {Kamionkowski}},\ and\
  \bibinfo {author} {\bibfnamefont {L.}~\bibnamefont {Dai}},\ }\bibfield
  {title} {\bibinfo {title} {{Pulsar-timing arrays, astrometry, and
  gravitational waves}},\ }\href {https://doi.org/10.1103/PhysRevD.99.063002}
  {\bibfield  {journal} {\bibinfo  {journal} {Phys. Rev. D}\ }\textbf {\bibinfo
  {volume} {99}},\ \bibinfo {pages} {063002} (\bibinfo {year} {2019})},\
  \Eprint {https://arxiv.org/abs/1810.02369} {arXiv:1810.02369 [astro-ph.CO]}
  \BibitemShut {NoStop}%
\bibitem [{\citenamefont {Ravi}\ \emph {et~al.}(2012)\citenamefont {Ravi},
  \citenamefont {Wyithe}, \citenamefont {Hobbs}, \citenamefont {Shannon},
  \citenamefont {Manchester}, \citenamefont {Yardley},\ and\ \citenamefont
  {Keith}}]{Ravi:2012bz}%
  \BibitemOpen
  \bibfield  {author} {\bibinfo {author} {\bibfnamefont {V.}~\bibnamefont
  {Ravi}}, \bibinfo {author} {\bibfnamefont {J.~S.~B.}\ \bibnamefont {Wyithe}},
  \bibinfo {author} {\bibfnamefont {G.}~\bibnamefont {Hobbs}}, \bibinfo
  {author} {\bibfnamefont {R.~M.}\ \bibnamefont {Shannon}}, \bibinfo {author}
  {\bibfnamefont {R.~N.}\ \bibnamefont {Manchester}}, \bibinfo {author}
  {\bibfnamefont {D.~R.~B.}\ \bibnamefont {Yardley}},\ and\ \bibinfo {author}
  {\bibfnamefont {M.~J.}\ \bibnamefont {Keith}},\ }\bibfield  {title} {\bibinfo
  {title} {{Does a 'stochastic' background of gravitational waves exist in the
  pulsar timing band?}},\ }\href {https://doi.org/10.1088/0004-637X/761/2/84}
  {\bibfield  {journal} {\bibinfo  {journal} {Astrophys. J.}\ }\textbf
  {\bibinfo {volume} {761}},\ \bibinfo {pages} {84} (\bibinfo {year} {2012})},\
  \Eprint {https://arxiv.org/abs/1210.3854} {arXiv:1210.3854 [astro-ph.CO]}
  \BibitemShut {NoStop}%
\bibitem [{\citenamefont {Cornish}\ and\ \citenamefont
  {Sesana}(2013)}]{Cornish:2013aba}%
  \BibitemOpen
  \bibfield  {author} {\bibinfo {author} {\bibfnamefont {N.~J.}\ \bibnamefont
  {Cornish}}\ and\ \bibinfo {author} {\bibfnamefont {A.}~\bibnamefont
  {Sesana}},\ }\bibfield  {title} {\bibinfo {title} {{Pulsar Timing Array
  Analysis for Black Hole Backgrounds}},\ }\href
  {https://doi.org/10.1088/0264-9381/30/22/224005} {\bibfield  {journal}
  {\bibinfo  {journal} {Class. Quant. Grav.}\ }\textbf {\bibinfo {volume}
  {30}},\ \bibinfo {pages} {224005} (\bibinfo {year} {2013})},\ \Eprint
  {https://arxiv.org/abs/1305.0326} {arXiv:1305.0326 [gr-qc]} \BibitemShut
  {NoStop}%
\bibitem [{\citenamefont {Sesana}\ \emph {et~al.}(2008)\citenamefont {Sesana},
  \citenamefont {Vecchio},\ and\ \citenamefont {Colacino}}]{Sesana:2008mz}%
  \BibitemOpen
  \bibfield  {author} {\bibinfo {author} {\bibfnamefont {A.}~\bibnamefont
  {Sesana}}, \bibinfo {author} {\bibfnamefont {A.}~\bibnamefont {Vecchio}},\
  and\ \bibinfo {author} {\bibfnamefont {C.~N.}\ \bibnamefont {Colacino}},\
  }\bibfield  {title} {\bibinfo {title} {{The stochastic gravitational-wave
  background from massive black hole binary systems: implications for
  observations with Pulsar Timing Arrays}},\ }\href
  {https://doi.org/10.1111/j.1365-2966.2008.13682.x} {\bibfield  {journal}
  {\bibinfo  {journal} {Mon. Not. Roy. Astron. Soc.}\ }\textbf {\bibinfo
  {volume} {390}},\ \bibinfo {pages} {192} (\bibinfo {year} {2008})},\ \Eprint
  {https://arxiv.org/abs/0804.4476} {arXiv:0804.4476 [astro-ph]} \BibitemShut
  {NoStop}%
\bibitem [{\citenamefont {Gair}\ \emph {et~al.}(2014)\citenamefont {Gair},
  \citenamefont {Romano}, \citenamefont {Taylor},\ and\ \citenamefont
  {Mingarelli}}]{Gair:2014rwa}%
  \BibitemOpen
  \bibfield  {author} {\bibinfo {author} {\bibfnamefont {J.}~\bibnamefont
  {Gair}}, \bibinfo {author} {\bibfnamefont {J.~D.}\ \bibnamefont {Romano}},
  \bibinfo {author} {\bibfnamefont {S.}~\bibnamefont {Taylor}},\ and\ \bibinfo
  {author} {\bibfnamefont {C.~M.~F.}\ \bibnamefont {Mingarelli}},\ }\bibfield
  {title} {\bibinfo {title} {{Mapping gravitational-wave backgrounds using
  methods from CMB analysis: Application to pulsar timing arrays}},\ }\href
  {https://doi.org/10.1103/PhysRevD.90.082001} {\bibfield  {journal} {\bibinfo
  {journal} {Phys. Rev. D}\ }\textbf {\bibinfo {volume} {90}},\ \bibinfo
  {pages} {082001} (\bibinfo {year} {2014})},\ \Eprint
  {https://arxiv.org/abs/1406.4664} {arXiv:1406.4664 [gr-qc]} \BibitemShut
  {NoStop}%
\bibitem [{\citenamefont {Taylor}\ and\ \citenamefont
  {Gair}(2013)}]{Taylor:2013esa}%
  \BibitemOpen
  \bibfield  {author} {\bibinfo {author} {\bibfnamefont {S.~R.}\ \bibnamefont
  {Taylor}}\ and\ \bibinfo {author} {\bibfnamefont {J.~R.}\ \bibnamefont
  {Gair}},\ }\bibfield  {title} {\bibinfo {title} {{Searching For Anisotropic
  Gravitational-wave Backgrounds Using Pulsar Timing Arrays}},\ }\href
  {https://doi.org/10.1103/PhysRevD.88.084001} {\bibfield  {journal} {\bibinfo
  {journal} {Phys. Rev. D}\ }\textbf {\bibinfo {volume} {88}},\ \bibinfo
  {pages} {084001} (\bibinfo {year} {2013})},\ \Eprint
  {https://arxiv.org/abs/1306.5395} {arXiv:1306.5395 [gr-qc]} \BibitemShut
  {NoStop}%
\bibitem [{\citenamefont {Hotinli}\ \emph {et~al.}(2019)\citenamefont
  {Hotinli}, \citenamefont {Kamionkowski},\ and\ \citenamefont
  {Jaffe}}]{Hotinli:2019tpc}%
  \BibitemOpen
  \bibfield  {author} {\bibinfo {author} {\bibfnamefont {S.~C.}\ \bibnamefont
  {Hotinli}}, \bibinfo {author} {\bibfnamefont {M.}~\bibnamefont
  {Kamionkowski}},\ and\ \bibinfo {author} {\bibfnamefont {A.~H.}\ \bibnamefont
  {Jaffe}},\ }\bibfield  {title} {\bibinfo {title} {{The search for anisotropy
  in the gravitational-wave background with pulsar-timing arrays}},\ }\href
  {https://doi.org/10.21105/astro.1904.05348} {\bibfield  {journal} {\bibinfo
  {journal} {Open J. Astrophys.}\ }\textbf {\bibinfo {volume} {2}},\ \bibinfo
  {pages} {8} (\bibinfo {year} {2019})},\ \Eprint
  {https://arxiv.org/abs/1904.05348} {arXiv:1904.05348 [astro-ph.CO]}
  \BibitemShut {NoStop}%
\bibitem [{\citenamefont {Ali-Ha\"\i{}moud}\ \emph {et~al.}(2021)\citenamefont
  {Ali-Ha\"\i{}moud}, \citenamefont {Smith},\ and\ \citenamefont
  {Mingarelli}}]{Ali-Haimoud:2020iyz}%
  \BibitemOpen
  \bibfield  {author} {\bibinfo {author} {\bibfnamefont {Y.}~\bibnamefont
  {Ali-Ha\"\i{}moud}}, \bibinfo {author} {\bibfnamefont {T.~L.}\ \bibnamefont
  {Smith}},\ and\ \bibinfo {author} {\bibfnamefont {C.~M.~F.}\ \bibnamefont
  {Mingarelli}},\ }\bibfield  {title} {\bibinfo {title} {{Insights into
  searches for anisotropies in the nanohertz gravitational-wave background}},\
  }\href {https://doi.org/10.1103/PhysRevD.103.042009} {\bibfield  {journal}
  {\bibinfo  {journal} {Phys. Rev. D}\ }\textbf {\bibinfo {volume} {103}},\
  \bibinfo {pages} {042009} (\bibinfo {year} {2021})},\ \Eprint
  {https://arxiv.org/abs/2010.13958} {arXiv:2010.13958 [gr-qc]} \BibitemShut
  {NoStop}%
\bibitem [{\citenamefont {Ali-Ha\"\i{}moud}\ \emph {et~al.}(2020)\citenamefont
  {Ali-Ha\"\i{}moud}, \citenamefont {Smith},\ and\ \citenamefont
  {Mingarelli}}]{Ali-Haimoud:2020ozu}%
  \BibitemOpen
  \bibfield  {author} {\bibinfo {author} {\bibfnamefont {Y.}~\bibnamefont
  {Ali-Ha\"\i{}moud}}, \bibinfo {author} {\bibfnamefont {T.~L.}\ \bibnamefont
  {Smith}},\ and\ \bibinfo {author} {\bibfnamefont {C.~M.~F.}\ \bibnamefont
  {Mingarelli}},\ }\bibfield  {title} {\bibinfo {title} {{Fisher formalism for
  anisotropic gravitational-wave background searches with pulsar timing
  arrays}},\ }\href {https://doi.org/10.1103/PhysRevD.102.122005} {\bibfield
  {journal} {\bibinfo  {journal} {Phys. Rev. D}\ }\textbf {\bibinfo {volume}
  {102}},\ \bibinfo {pages} {122005} (\bibinfo {year} {2020})},\ \Eprint
  {https://arxiv.org/abs/2006.14570} {arXiv:2006.14570 [gr-qc]} \BibitemShut
  {NoStop}%
\bibitem [{\citenamefont {Pol}\ \emph {et~al.}(2022)\citenamefont {Pol},
  \citenamefont {Taylor},\ and\ \citenamefont {Romano}}]{Pol:2022sjn}%
  \BibitemOpen
  \bibfield  {author} {\bibinfo {author} {\bibfnamefont {N.}~\bibnamefont
  {Pol}}, \bibinfo {author} {\bibfnamefont {S.~R.}\ \bibnamefont {Taylor}},\
  and\ \bibinfo {author} {\bibfnamefont {J.~D.}\ \bibnamefont {Romano}},\
  }\bibfield  {title} {\bibinfo {title} {{Forecasting Pulsar Timing Array
  Sensitivity to Anisotropy in the Stochastic Gravitational Wave Background}},\
  }\href {https://doi.org/10.3847/1538-4357/ac9836} {\bibfield  {journal}
  {\bibinfo  {journal} {Astrophys. J.}\ }\textbf {\bibinfo {volume} {940}},\
  \bibinfo {pages} {173} (\bibinfo {year} {2022})},\ \Eprint
  {https://arxiv.org/abs/2206.09936} {arXiv:2206.09936 [astro-ph.HE]}
  \BibitemShut {NoStop}%
\bibitem [{\citenamefont {Taylor}\ \emph {et~al.}(2015)\citenamefont {Taylor}
  \emph {et~al.}}]{Taylor:2015udp}%
  \BibitemOpen
  \bibfield  {author} {\bibinfo {author} {\bibfnamefont {S.~R.}\ \bibnamefont
  {Taylor}} \emph {et~al.},\ }\bibfield  {title} {\bibinfo {title} {{Limits on
  anisotropy in the nanohertz stochastic gravitational-wave background}},\
  }\href {https://doi.org/10.1103/PhysRevLett.115.041101} {\bibfield  {journal}
  {\bibinfo  {journal} {Phys. Rev. Lett.}\ }\textbf {\bibinfo {volume} {115}},\
  \bibinfo {pages} {041101} (\bibinfo {year} {2015})},\ \Eprint
  {https://arxiv.org/abs/1506.08817} {arXiv:1506.08817 [astro-ph.HE]}
  \BibitemShut {NoStop}%
\bibitem [{\citenamefont {Agazie}\ \emph
  {et~al.}(2023{\natexlab{b}})\citenamefont {Agazie} \emph
  {et~al.}}]{NANOGrav:2023tcn}%
  \BibitemOpen
  \bibfield  {author} {\bibinfo {author} {\bibfnamefont {G.}~\bibnamefont
  {Agazie}} \emph {et~al.} (\bibinfo {collaboration} {NANOGrav}),\ }\bibfield
  {title} {\bibinfo {title} {{The NANOGrav 15 yr Data Set: Search for
  Anisotropy in the Gravitational-wave Background}},\ }\href
  {https://doi.org/10.3847/2041-8213/acf4fd} {\bibfield  {journal} {\bibinfo
  {journal} {Astrophys. J. Lett.}\ }\textbf {\bibinfo {volume} {956}},\
  \bibinfo {pages} {L3} (\bibinfo {year} {2023}{\natexlab{b}})},\ \Eprint
  {https://arxiv.org/abs/2306.16221} {arXiv:2306.16221 [astro-ph.HE]}
  \BibitemShut {NoStop}%
\bibitem [{\citenamefont {Kato}\ and\ \citenamefont
  {Soda}(2016)}]{Kato:2015bye}%
  \BibitemOpen
  \bibfield  {author} {\bibinfo {author} {\bibfnamefont {R.}~\bibnamefont
  {Kato}}\ and\ \bibinfo {author} {\bibfnamefont {J.}~\bibnamefont {Soda}},\
  }\bibfield  {title} {\bibinfo {title} {{Probing circular polarization in
  stochastic gravitational wave background with pulsar timing arrays}},\ }\href
  {https://doi.org/10.1103/PhysRevD.93.062003} {\bibfield  {journal} {\bibinfo
  {journal} {Phys. Rev. D}\ }\textbf {\bibinfo {volume} {93}},\ \bibinfo
  {pages} {062003} (\bibinfo {year} {2016})},\ \Eprint
  {https://arxiv.org/abs/1512.09139} {arXiv:1512.09139 [gr-qc]} \BibitemShut
  {NoStop}%
\bibitem [{\citenamefont {Belgacem}\ and\ \citenamefont
  {Kamionkowski}(2020)}]{Belgacem:2020nda}%
  \BibitemOpen
  \bibfield  {author} {\bibinfo {author} {\bibfnamefont {E.}~\bibnamefont
  {Belgacem}}\ and\ \bibinfo {author} {\bibfnamefont {M.}~\bibnamefont
  {Kamionkowski}},\ }\bibfield  {title} {\bibinfo {title} {{Chirality of the
  gravitational-wave background and pulsar-timing arrays}},\ }\href
  {https://doi.org/10.1103/PhysRevD.102.023004} {\bibfield  {journal} {\bibinfo
   {journal} {Phys. Rev. D}\ }\textbf {\bibinfo {volume} {102}},\ \bibinfo
  {pages} {023004} (\bibinfo {year} {2020})},\ \Eprint
  {https://arxiv.org/abs/2004.05480} {arXiv:2004.05480 [astro-ph.CO]}
  \BibitemShut {NoStop}%
\bibitem [{\citenamefont {Sato-Polito}\ and\ \citenamefont
  {Kamionkowski}(2022)}]{Sato-Polito:2021efu}%
  \BibitemOpen
  \bibfield  {author} {\bibinfo {author} {\bibfnamefont {G.}~\bibnamefont
  {Sato-Polito}}\ and\ \bibinfo {author} {\bibfnamefont {M.}~\bibnamefont
  {Kamionkowski}},\ }\bibfield  {title} {\bibinfo {title} {{Pulsar-timing
  measurement of the circular polarization of the stochastic gravitational-wave
  background}},\ }\href {https://doi.org/10.1103/PhysRevD.106.023004}
  {\bibfield  {journal} {\bibinfo  {journal} {Phys. Rev. D}\ }\textbf {\bibinfo
  {volume} {106}},\ \bibinfo {pages} {023004} (\bibinfo {year} {2022})},\
  \Eprint {https://arxiv.org/abs/2111.05867} {arXiv:2111.05867 [astro-ph.CO]}
  \BibitemShut {NoStop}%
\bibitem [{\citenamefont {Chu}\ \emph {et~al.}(2021)\citenamefont {Chu},
  \citenamefont {Liu},\ and\ \citenamefont {Ng}}]{Chu:2021krj}%
  \BibitemOpen
  \bibfield  {author} {\bibinfo {author} {\bibfnamefont {Y.-K.}\ \bibnamefont
  {Chu}}, \bibinfo {author} {\bibfnamefont {G.-C.}\ \bibnamefont {Liu}},\ and\
  \bibinfo {author} {\bibfnamefont {K.-W.}\ \bibnamefont {Ng}},\ }\bibfield
  {title} {\bibinfo {title} {{Observation of a polarized stochastic
  gravitational-wave background in pulsar-timing-array experiments}},\ }\href
  {https://doi.org/10.1103/PhysRevD.104.124018} {\bibfield  {journal} {\bibinfo
   {journal} {Phys. Rev. D}\ }\textbf {\bibinfo {volume} {104}},\ \bibinfo
  {pages} {124018} (\bibinfo {year} {2021})},\ \Eprint
  {https://arxiv.org/abs/2107.00536} {arXiv:2107.00536 [gr-qc]} \BibitemShut
  {NoStop}%
\bibitem [{\citenamefont {Liu}\ and\ \citenamefont {Ng}(2022)}]{Liu:2022skj}%
  \BibitemOpen
  \bibfield  {author} {\bibinfo {author} {\bibfnamefont {G.-C.}\ \bibnamefont
  {Liu}}\ and\ \bibinfo {author} {\bibfnamefont {K.-W.}\ \bibnamefont {Ng}},\
  }\bibfield  {title} {\bibinfo {title} {{Timing-residual power spectrum of a
  polarized stochastic gravitational-wave background in pulsar-timing-array
  observation}},\ }\href {https://doi.org/10.1103/PhysRevD.106.064004}
  {\bibfield  {journal} {\bibinfo  {journal} {Phys. Rev. D}\ }\textbf {\bibinfo
  {volume} {106}},\ \bibinfo {pages} {064004} (\bibinfo {year} {2022})},\
  \Eprint {https://arxiv.org/abs/2201.06767} {arXiv:2201.06767 [gr-qc]}
  \BibitemShut {NoStop}%
\bibitem [{\citenamefont {Hajian}\ and\ \citenamefont
  {Souradeep}(2003)}]{Hajian:2003qq}%
  \BibitemOpen
  \bibfield  {author} {\bibinfo {author} {\bibfnamefont {A.}~\bibnamefont
  {Hajian}}\ and\ \bibinfo {author} {\bibfnamefont {T.}~\bibnamefont
  {Souradeep}},\ }\bibfield  {title} {\bibinfo {title} {{Measuring statistical
  isotropy of the CMB anisotropy}},\ }\href {https://doi.org/10.1086/379757}
  {\bibfield  {journal} {\bibinfo  {journal} {Astrophys. J. Lett.}\ }\textbf
  {\bibinfo {volume} {597}},\ \bibinfo {pages} {L5} (\bibinfo {year} {2003})},\
  \Eprint {https://arxiv.org/abs/astro-ph/0308001} {arXiv:astro-ph/0308001}
  \BibitemShut {NoStop}%
\bibitem [{\citenamefont {{Hajian}}\ and\ \citenamefont
  {{Souradeep}}(2004)}]{Hajian:2005jh}%
  \BibitemOpen
  \bibfield  {author} {\bibinfo {author} {\bibfnamefont {A.}~\bibnamefont
  {{Hajian}}}\ and\ \bibinfo {author} {\bibfnamefont {T.}~\bibnamefont
  {{Souradeep}}},\ }\bibfield  {title} {\bibinfo {title} {{The Cosmic Microwave
  Background Bipolar Power Spectrum: Basic Formalism and Applications}},\
  }\href {https://doi.org/10.48550/arXiv.astro-ph/0501001} {\bibfield
  {journal} {\bibinfo  {journal} {arXiv e-prints}\ ,\ \bibinfo {eid}
  {astro-ph/0501001}} (\bibinfo {year} {2004})},\ \Eprint
  {https://arxiv.org/abs/astro-ph/0501001} {arXiv:astro-ph/0501001 [astro-ph]}
  \BibitemShut {NoStop}%
\bibitem [{\citenamefont {Joshi}\ \emph {et~al.}(2010)\citenamefont {Joshi},
  \citenamefont {Jhingan}, \citenamefont {Souradeep},\ and\ \citenamefont
  {Hajian}}]{Joshi:2009mj}%
  \BibitemOpen
  \bibfield  {author} {\bibinfo {author} {\bibfnamefont {N.}~\bibnamefont
  {Joshi}}, \bibinfo {author} {\bibfnamefont {S.}~\bibnamefont {Jhingan}},
  \bibinfo {author} {\bibfnamefont {T.}~\bibnamefont {Souradeep}},\ and\
  \bibinfo {author} {\bibfnamefont {A.}~\bibnamefont {Hajian}},\ }\bibfield
  {title} {\bibinfo {title} {{Bipolar Harmonic encoding of CMB correlation
  patterns}},\ }\href {https://doi.org/10.1103/PhysRevD.81.083012} {\bibfield
  {journal} {\bibinfo  {journal} {Phys. Rev. D}\ }\textbf {\bibinfo {volume}
  {81}},\ \bibinfo {pages} {083012} (\bibinfo {year} {2010})},\ \Eprint
  {https://arxiv.org/abs/0912.3217} {arXiv:0912.3217 [astro-ph.CO]}
  \BibitemShut {NoStop}%
\bibitem [{\citenamefont {Book}\ \emph {et~al.}(2012)\citenamefont {Book},
  \citenamefont {Kamionkowski},\ and\ \citenamefont {Souradeep}}]{Book:2011na}%
  \BibitemOpen
  \bibfield  {author} {\bibinfo {author} {\bibfnamefont {L.~G.}\ \bibnamefont
  {Book}}, \bibinfo {author} {\bibfnamefont {M.}~\bibnamefont {Kamionkowski}},\
  and\ \bibinfo {author} {\bibfnamefont {T.}~\bibnamefont {Souradeep}},\
  }\bibfield  {title} {\bibinfo {title} {{Odd-Parity Bipolar Spherical
  Harmonics}},\ }\href {https://doi.org/10.1103/PhysRevD.85.023010} {\bibfield
  {journal} {\bibinfo  {journal} {Phys. Rev. D}\ }\textbf {\bibinfo {volume}
  {85}},\ \bibinfo {pages} {023010} (\bibinfo {year} {2012})},\ \Eprint
  {https://arxiv.org/abs/1109.2910} {arXiv:1109.2910 [astro-ph.CO]}
  \BibitemShut {NoStop}%
\bibitem [{\citenamefont {Anil~Kumar}\ and\ \citenamefont
  {Kamionkowski}(2023)}]{AnilKumar:2023yfw}%
  \BibitemOpen
  \bibfield  {author} {\bibinfo {author} {\bibfnamefont {N.}~\bibnamefont
  {Anil~Kumar}}\ and\ \bibinfo {author} {\bibfnamefont {M.}~\bibnamefont
  {Kamionkowski}},\ }\bibfield  {title} {\bibinfo {title} {{All the Pretty
  Overlap Reduction Functions}},\ }\href@noop {} {\  (\bibinfo {year}
  {2023})},\ \Eprint {https://arxiv.org/abs/2311.14159} {arXiv:2311.14159
  [astro-ph.CO]} \BibitemShut {NoStop}%
\bibitem [{\citenamefont {Bernardo}\ \emph {et~al.}(2023)\citenamefont
  {Bernardo}, \citenamefont {Liu},\ and\ \citenamefont
  {Ng}}]{Bernardo:2023jhs}%
  \BibitemOpen
  \bibfield  {author} {\bibinfo {author} {\bibfnamefont {R.~C.}\ \bibnamefont
  {Bernardo}}, \bibinfo {author} {\bibfnamefont {G.-C.}\ \bibnamefont {Liu}},\
  and\ \bibinfo {author} {\bibfnamefont {K.-W.}\ \bibnamefont {Ng}},\
  }\href@noop {} {\bibinfo {title} {{Correlations for an anisotropic polarized
  stochastic gravitational wave background in pulsar timing arrays}}} (\bibinfo
  {year} {2023}),\ \Eprint {https://arxiv.org/abs/2312.03383} {arXiv:2312.03383
  [gr-qc]} \BibitemShut {NoStop}%
\bibitem [{\citenamefont {Mingarelli}\ \emph {et~al.}(2013)\citenamefont
  {Mingarelli}, \citenamefont {Sidery}, \citenamefont {Mandel},\ and\
  \citenamefont {Vecchio}}]{Mingarelli:2013dsa}%
  \BibitemOpen
  \bibfield  {author} {\bibinfo {author} {\bibfnamefont {C.~M.~F.}\
  \bibnamefont {Mingarelli}}, \bibinfo {author} {\bibfnamefont
  {T.}~\bibnamefont {Sidery}}, \bibinfo {author} {\bibfnamefont
  {I.}~\bibnamefont {Mandel}},\ and\ \bibinfo {author} {\bibfnamefont
  {A.}~\bibnamefont {Vecchio}},\ }\bibfield  {title} {\bibinfo {title}
  {{Characterizing gravitational wave stochastic background anisotropy with
  pulsar timing arrays}},\ }\href {https://doi.org/10.1103/PhysRevD.88.062005}
  {\bibfield  {journal} {\bibinfo  {journal} {Phys. Rev. D}\ }\textbf {\bibinfo
  {volume} {88}},\ \bibinfo {pages} {062005} (\bibinfo {year} {2013})},\
  \Eprint {https://arxiv.org/abs/1306.5394} {arXiv:1306.5394 [astro-ph.HE]}
  \BibitemShut {NoStop}%
\bibitem [{\citenamefont {Sato-Polito}\ and\ \citenamefont
  {Kamionkowski}(2023)}]{Sato-Polito:2023spo}%
  \BibitemOpen
  \bibfield  {author} {\bibinfo {author} {\bibfnamefont {G.}~\bibnamefont
  {Sato-Polito}}\ and\ \bibinfo {author} {\bibfnamefont {M.}~\bibnamefont
  {Kamionkowski}},\ }\bibfield  {title} {\bibinfo {title} {{Exploring the
  spectrum of stochastic gravitational-wave anisotropies with pulsar timing
  arrays}},\ }\href@noop {} {\  (\bibinfo {year} {2023})},\ \Eprint
  {https://arxiv.org/abs/2305.05690} {arXiv:2305.05690 [astro-ph.CO]}
  \BibitemShut {NoStop}%
\bibitem [{\citenamefont {Gardiner}\ \emph {et~al.}(2023)\citenamefont
  {Gardiner}, \citenamefont {Kelley}, \citenamefont {Lemke},\ and\
  \citenamefont {Mitridate}}]{Gardiner:2023zzr}%
  \BibitemOpen
  \bibfield  {author} {\bibinfo {author} {\bibfnamefont {E.~C.}\ \bibnamefont
  {Gardiner}}, \bibinfo {author} {\bibfnamefont {L.~Z.}\ \bibnamefont
  {Kelley}}, \bibinfo {author} {\bibfnamefont {A.-M.}\ \bibnamefont {Lemke}},\
  and\ \bibinfo {author} {\bibfnamefont {A.}~\bibnamefont {Mitridate}},\
  }\bibfield  {title} {\bibinfo {title} {{Beyond the Background: Gravitational
  Wave Anisotropy and Continuous Waves from Supermassive Black Hole
  Binaries}},\ }\href@noop {} {\  (\bibinfo {year} {2023})},\ \Eprint
  {https://arxiv.org/abs/2309.07227} {arXiv:2309.07227 [astro-ph.HE]}
  \BibitemShut {NoStop}%
\bibitem [{\citenamefont {{Sazhin}}(1978)}]{1978SvA....22...36S}%
  \BibitemOpen
  \bibfield  {author} {\bibinfo {author} {\bibfnamefont {M.~V.}\ \bibnamefont
  {{Sazhin}}},\ }\bibfield  {title} {\bibinfo {title} {{Opportunities for
  detecting ultralong gravitational waves}},\ }\href@noop {} {\bibfield
  {journal} {\bibinfo  {journal} {Sov.\ Astron.}\ }\textbf {\bibinfo {volume}
  {22}},\ \bibinfo {pages} {36} (\bibinfo {year} {1978})}\BibitemShut {NoStop}%
\bibitem [{\citenamefont {{Detweiler}}(1979)}]{1979ApJ...234.1100D}%
  \BibitemOpen
  \bibfield  {author} {\bibinfo {author} {\bibfnamefont {S.}~\bibnamefont
  {{Detweiler}}},\ }\bibfield  {title} {\bibinfo {title} {{Pulsar timing
  measurements and the search for gravitational waves}},\ }\href
  {https://doi.org/10.1086/157593} {\bibfield  {journal} {\bibinfo  {journal}
  {\apj}\ }\textbf {\bibinfo {volume} {234}},\ \bibinfo {pages} {1100}
  (\bibinfo {year} {1979})}\BibitemShut {NoStop}%
\bibitem [{\citenamefont {Conneely}\ \emph {et~al.}(2019)\citenamefont
  {Conneely}, \citenamefont {Jaffe},\ and\ \citenamefont
  {Mingarelli}}]{Conneely:2018wis}%
  \BibitemOpen
  \bibfield  {author} {\bibinfo {author} {\bibfnamefont {C.}~\bibnamefont
  {Conneely}}, \bibinfo {author} {\bibfnamefont {A.~H.}\ \bibnamefont
  {Jaffe}},\ and\ \bibinfo {author} {\bibfnamefont {C.~M.~F.}\ \bibnamefont
  {Mingarelli}},\ }\bibfield  {title} {\bibinfo {title} {{On the Amplitude and
  Stokes Parameters of a Stochastic Gravitational-Wave Background}},\ }\href
  {https://doi.org/10.1093/mnras/stz1022} {\bibfield  {journal} {\bibinfo
  {journal} {Mon. Not. Roy. Astron. Soc.}\ }\textbf {\bibinfo {volume} {487}},\
  \bibinfo {pages} {562} (\bibinfo {year} {2019})},\ \Eprint
  {https://arxiv.org/abs/1808.05920} {arXiv:1808.05920 [astro-ph.CO]}
  \BibitemShut {NoStop}%
\bibitem [{\citenamefont {Roebber}\ and\ \citenamefont
  {Holder}(2017)}]{Roebber:2016jzl}%
  \BibitemOpen
  \bibfield  {author} {\bibinfo {author} {\bibfnamefont {E.}~\bibnamefont
  {Roebber}}\ and\ \bibinfo {author} {\bibfnamefont {G.}~\bibnamefont
  {Holder}},\ }\bibfield  {title} {\bibinfo {title} {{Harmonic space analysis
  of pulsar timing array redshift maps}},\ }\href
  {https://doi.org/10.3847/1538-4357/835/1/21} {\bibfield  {journal} {\bibinfo
  {journal} {Astrophys. J.}\ }\textbf {\bibinfo {volume} {835}},\ \bibinfo
  {pages} {21} (\bibinfo {year} {2017})},\ \Eprint
  {https://arxiv.org/abs/1609.06758} {arXiv:1609.06758 [astro-ph.CO]}
  \BibitemShut {NoStop}%
\bibitem [{\citenamefont {{Mingarelli}}\ and\ \citenamefont
  {{Mingarelli}}(2018)}]{2018JPhCo...2j5002M}%
  \BibitemOpen
  \bibfield  {author} {\bibinfo {author} {\bibfnamefont {C.~M.~F.}\
  \bibnamefont {{Mingarelli}}}\ and\ \bibinfo {author} {\bibfnamefont {A.~B.}\
  \bibnamefont {{Mingarelli}}},\ }\bibfield  {title} {\bibinfo {title}
  {{Proving the short-wavelength approximation in Pulsar Timing Array
  gravitational-wave background searches}},\ }\href
  {https://doi.org/10.1088/2399-6528/aae06d} {\bibfield  {journal} {\bibinfo
  {journal} {Journal of Physics Communications}\ }\textbf {\bibinfo {volume}
  {2}},\ \bibinfo {eid} {105002} (\bibinfo {year} {2018})},\ \Eprint
  {https://arxiv.org/abs/1806.06979} {arXiv:1806.06979 [astro-ph.IM]}
  \BibitemShut {NoStop}%
\bibitem [{\citenamefont {Khersonskii}\ \emph {et~al.}(1988)\citenamefont
  {Khersonskii}, \citenamefont {Moskalev},\ and\ \citenamefont
  {Varshalovich}}]{Khersonskii:1988krb}%
  \BibitemOpen
  \bibfield  {author} {\bibinfo {author} {\bibfnamefont {V.~K.}\ \bibnamefont
  {Khersonskii}}, \bibinfo {author} {\bibfnamefont {A.~N.}\ \bibnamefont
  {Moskalev}},\ and\ \bibinfo {author} {\bibfnamefont {D.~A.}\ \bibnamefont
  {Varshalovich}},\ }\href {https://doi.org/10.1142/0270} {\emph {\bibinfo
  {title} {{Quantum Theory Of Angular Momentum}}}}\ (\bibinfo  {publisher}
  {World Scientific Publishing Company},\ \bibinfo {year} {1988})\BibitemShut
  {NoStop}%
\bibitem [{\citenamefont {Gair}\ \emph {et~al.}(2015)\citenamefont {Gair},
  \citenamefont {Romano},\ and\ \citenamefont {Taylor}}]{Gair:2015hra}%
  \BibitemOpen
  \bibfield  {author} {\bibinfo {author} {\bibfnamefont {J.~R.}\ \bibnamefont
  {Gair}}, \bibinfo {author} {\bibfnamefont {J.~D.}\ \bibnamefont {Romano}},\
  and\ \bibinfo {author} {\bibfnamefont {S.~R.}\ \bibnamefont {Taylor}},\
  }\bibfield  {title} {\bibinfo {title} {{Mapping gravitational-wave
  backgrounds of arbitrary polarisation using pulsar timing arrays}},\ }\href
  {https://doi.org/10.1103/PhysRevD.92.102003} {\bibfield  {journal} {\bibinfo
  {journal} {Phys. Rev. D}\ }\textbf {\bibinfo {volume} {92}},\ \bibinfo
  {pages} {102003} (\bibinfo {year} {2015})},\ \Eprint
  {https://arxiv.org/abs/1506.08668} {arXiv:1506.08668 [gr-qc]} \BibitemShut
  {NoStop}%
\bibitem [{\citenamefont {Ng}(2017)}]{Ng:2017djg}%
  \BibitemOpen
  \bibfield  {author} {\bibinfo {author} {\bibfnamefont {C.}~\bibnamefont {Ng}}
  (\bibinfo {collaboration} {CHIME Pulsar}),\ }\bibfield  {title} {\bibinfo
  {title} {{Pulsar science with the CHIME telescope}},\ }\href
  {https://doi.org/10.1017/S1743921317010638} {\bibfield  {journal} {\bibinfo
  {journal} {IAU Symp.}\ }\textbf {\bibinfo {volume} {337}},\ \bibinfo {pages}
  {179} (\bibinfo {year} {2017})},\ \Eprint {https://arxiv.org/abs/1711.02104}
  {arXiv:1711.02104 [astro-ph.IM]} \BibitemShut {NoStop}%
\bibitem [{\citenamefont {Bailes}\ \emph {et~al.}(2018)\citenamefont {Bailes}
  \emph {et~al.}}]{Bailes:2018azh}%
  \BibitemOpen
  \bibfield  {author} {\bibinfo {author} {\bibfnamefont {M.}~\bibnamefont
  {Bailes}} \emph {et~al.},\ }\bibfield  {title} {\bibinfo {title} {{MeerTime -
  the MeerKAT Key Science Program on Pulsar Timing}},\ }\href
  {https://doi.org/10.22323/1.277.0011} {\bibfield  {journal} {\bibinfo
  {journal} {PoS}\ }\textbf {\bibinfo {volume} {MeerKAT2016}},\ \bibinfo
  {pages} {011} (\bibinfo {year} {2018})},\ \Eprint
  {https://arxiv.org/abs/1803.07424} {arXiv:1803.07424 [astro-ph.IM]}
  \BibitemShut {NoStop}%
\end{thebibliography}%
\end{document}